\documentclass[
  journal=pasa,
  manuscript=research-paper, 
  year=2023,
  volume=,
]{cup-journal}

\usepackage{microtype,siunitx,booktabs}
\usepackage{graphicx}
\usepackage{amssymb}	
\usepackage{amsbsy}   
\usepackage{amsmath}
\usepackage{multirow}   
\usepackage{makecell}   
\usepackage{threeparttable}  
\usepackage{booktabs}   
\usepackage{natbib}
\usepackage{xcolor}
\usepackage{enumitem}

\newcommand{\lya}{Ly$\alpha$}
\newcommand{\lymana}{Lyman-$\alpha$}
\newcommand{\halpha}{H$\alpha$}

\newcommand{\HI}{H{\footnotesize I}}

\newcommand{\kms}{\,km\,s$^{-1}$}
\newcommand{\ugriz}{$u^* g^\prime r^\prime i^\prime z^\prime$}
\newcommand{\UGR}{$U_nG\cal{R}$}

\newcommand*{\ditto}{---\texttt{"}---}
\newcommand{\KMOS}{K$^{\mathrm{3D}}$}
\newcommand{\vobs}{${v}_{\mathrm{obs}}/2{\sigma }_{\mathrm{int}}$} 
\newcommand{\vrsO}{${v}_{\mathrm{rot}}/{\sigma}_{\mathrm{0}}$}
\newcommand{\vshear}{$v_{\mathrm{shear}}/{\sigma}_{\mathrm{mean}}$} 
\newcommand{\dblprime}{^{\prime\prime}}

\title[Kinematics and \lya\ in $z\sim$ 2--3 LBGs]{\lymana\ at Cosmic Noon II: The relationship between kinematics and \lymana\ in $z\sim$ 2--3 Lyman Break Galaxies}

\author{Garry Foran}
\affiliation{Centre for Astrophysics and Supercomputing, Swinburne University of Technology, PO Box 218, H29, Hawthorn, VIC 3122, Australia}
\alsoaffiliation{Australian Research Council Centre of Excellence for All-sky Astrophysics in 3 Dimensions (ASTRO-3D)}
\email{gforan@swin.edu.au}

\author{Jeff Cooke}
\affiliation{Centre for Astrophysics and Supercomputing, Swinburne University of Technology, PO Box 218, H29, Hawthorn, VIC 3122, Australia}
\alsoaffiliation{Australian Research Council Centre of Excellence for All-sky Astrophysics in 3 Dimensions (ASTRO-3D)}

\author{Emily Wisnioski}
\affiliation{Research School of Astronomy and Astrophysics, Australian National University, ACT 2611, Australia}
\alsoaffiliation{Australian Research Council Centre of Excellence for All-sky Astrophysics in 3 Dimensions (ASTRO-3D)}

\author{Naveen Reddy}
\affiliation{Department of Physics \& Astronomy, University of California, Riverside, 900 University Avenue, Riverside, CA 92521, USA}

\author{Charles Steidel}
\affiliation{Cahill Center for Astronomy and Astrophysics, California Institute of Technology, MS 249-17, Pasadena, CA, 92115, USA}

\doi{}

\received {dd Mmm YYYY}
\revised  {dd Mmm YYYY}
\accepted {dd Mmm YYYY}
\published{}


\begin{document}

\begin{abstract}
We report for the first time a relationship between galaxy kinematics and net \lymana\ equivalent width (net \lya\ EW) in star forming galaxies during the epoch of peak cosmic star formation.  Building on the previously reported broadband imaging segregation of \lya-emitting and \lya-absorbing Lyman-break galaxies (LBGs) at $z\sim2$ (Paper\,I in this series) and previously at $z\sim3$, we use the \lya\ spectral type classification method to study the relationship between net \lya\ EW and nebular emission-line kinematics in samples of $z\sim2$ and $z\sim3$ LBGs drawn from the literature for which matching rest-frame UV photometry, consistently measured net \lya\ EWs, and kinematic classifications from integral field unit spectroscopy are available.  we show that $z\sim2$ and $z\sim3$ LBGs segregate in colour--magnitude space according to their kinematic properties and \lymana\ spectral type, and conclude that LBGs with \lya\ dominant in absorption (aLBGs) are almost exclusively rotation-dominated (presumably disc-like) systems, and LBGs with \lya\ dominant in emission (eLBGs) characteristically have dispersion-dominated kinematics.  We quantify the  relationship between the strength of rotational dynamic support (as measured using \vobs\ and \vrsO) and net \lya\ EW for subsets of our kinematic sample where these data are available, and demonstrate the consistency of our result with other properties that scale with net \lya\ EW and kinematics.  Based on these findings, we suggest a method by which large samples of rotation- and dispersion-dominated galaxies might be selected using broadband imaging in as few as three filters and/or net \lya\ EW alone.  If confirmed with larger samples, application of this method will enable an understanding of galaxy kinematic behaviour over large scales in datasets from current and future large-area and all-sky photometric surveys that will select hundreds of millions of LBGs in redshift ranges from $z\sim 2-6$ across many hundreds to thousands of Mpc.  Finally, we speculate that the combination of our result linking net \lya\ EW and nebular emission-line kinematics with the known large-scale clustering behaviour of \lya-absorbing and \lya-emitting LBGs is evocative of an emergent bimodality of early galaxies that is consistent with a nascent morphology-density relation at $z\sim2-3$.
\end{abstract}

\section{INTRODUCTION} \label{sec:c4_intro}

Kinematics are a characteristic feature of the intrinsic galaxy population dichotomy that exists in the modern-day Universe \citep[see for example][]{Blanton2009}, and as such, are a key constraint for simulations that aim to understand the mechanisms by which galaxies evolve over cosmic time \citep[e.g.,][]{El-Badry2018, Hung2019, Pillepich2019, Meng2019, Vogelsberger2020, Dubois2020}.  This intimate relationship between galaxy kinematics and their environment is manifest in the well-studied Morphology--Density Relation \citep[MDR; e.g.,][]{Dressler1980, Postman2005, Houghton2015, Paulino2019, Sazonova2020}, and has been investigated for a range of kinematic types and environments out to intermediate redshifts up to $z\sim1.5$ \citep[e.g.,][]{Perez-Mart2017, Pelliccia2019, Bohm2020, Cole2020, Tiley2020}.

Developments in the capability and efficiency of optical and near-IR integral field unit (IFU) and slit spectrographs over the past fifteen years have enabled numerous observational campaigns to probe in exquisite detail the nebular emission-line kinematics of star-forming galaxies (SFGs) at redshifts that span the peak of the cosmic star formation rate density \citep[$z\approx 1-4$; ][]{Madau2014}.  The synthesis of these results has produced a picture of the high-redshift SFG population in which the majority ($\gtrsim 75$\%) of massive ($\mathrm{log}({M}_{\star }/{M}_{\odot }) \gtrsim 10$) SFGs appear to have assembled primitive discs with characteristically large ionised gas velocity dispersions and rotation-dominated kinematics \citep[e.g.,][]{Wisnioski2015, Wisnioski2019, Simons2016, Simons2017, FS2018}.  Lensed and other deep surveys targeting lower mass ($\mathrm{log}({M}_{\star }/{M}_{\odot })\lesssim 9.5$) SFGs report a significantly lower ($< 50\%$) rotation-dominated fraction, and a higher proportion of galaxies that are kinematically disordered (dispersion-dominated) or have kinematic structure and/or morphologies that identified them as mergers \citep[e.g.,][]{Gnerucci2011, Livermore2015, Leethochawalit2016, Mason2017, Turner2017, Girard2018}.  Additionally, the nebular emission-line kinematics of SFGs in the redshift range $z\sim2-3$, have been shown to correlate with a range of galactic physical properties including: stellar and dynamical mass; size and morphology; specific and total star formation rates; gas outflows and gas fraction; and nebular conditions including the degree and spatial distribution of metals (see for example \citet{Glazebrook2013} and references therein, and more recently, \citet{Wisnioski2015, Wisnioski2018, Wisnioski2019, Leethochawalit2016, Simons2016, Simons2017, Mason2017, Molina2017, Turner2017, Girard2018, FS2018, FS2019} and \citet{FS2020}).  

Despite these advances, the ability of currently available samples to inform the relationship between kinematics and large-scale structure at redshifts $z\gtrsim2$ is challenged by a range of factors including: small sample sizes; observational biases that are a function of survey depth and redshift range; specific sample selection criteria; the spatial and spectral resolution of the observations; and the diversity of kinematic analysis methods and classification criteria \citep[e.g.,][]{Glazebrook2013, Leethochawalit2016, di-Teodoro2016, Rodrigues2017, Simons2019, Hung2019}.  To better inform simulations that model the relationship between galactic kinematics and different formation and evolution pathways, and to facilitate the proper statistical study of relationships between the global kinematic properties of galaxy populations and large scale structure out to high redshifts, there is a need for much larger samples spanning a wider range of intrinsic and environmental galaxy properties, mapped over large scales, and to redshifts above $z\sim4$, at which redshifts IFU-based kinematic measurements are currently not possible.

Contemporaneously with progress in understanding the kinematics of SFGs at $z\gtrsim2$, rest-frame UV spectroscopic studies of star-forming UV-colour-selected Lyman break galaxies (LBGs) in the same redshift range have reported the sensitivity of \lymana\ (\lya) visibility to a wide range of galactic properties.  Due to the resonant character of the \lya\ transition (see M.~Dijkstra in \citet{SAAS-FE2019} for a comprehensive description), \lya\ transmission and spectral morphology are modulated by neutral gas properties such as  optical depth, covering fraction, dust content, and kinematics (see the review of \citet{Hayes2015} and more recently e.g., \citet{ Trainor2015, Trainor2019, Gronke2016a, Reddy2016, Steidel2018, Du2018, Remolina2019, Byrohl2020} and \citet{Pahl2020}).  In addition to these \HI\ gas properties which directly control the absorption and scattering of \lya\ during radiative transfer, it has been demonstrated for $z\sim2-4$ LBGs that larger \lymana\ transmission, or net \lymana\ equivalent width (net \lya\ EW), is associated with galaxies with bluer UV colours, lower metallicities, lower stellar masses, lower total UV luminosities, lower star formation rates, harder ionising field strengths, and more compact morphologies \citep[e.g.,][]{Shapley2003, Erb2006a, Law2007a, Steidel2010, Pentericci2010, Law2012c, Erb2016, Du2018, Marchi2019}.  

Deep observational surveys of the spatial redistribution of \lya\ into the circumgalactic medium (CGM) have established the apparent ubiquity of so-called `\lymana\ halos' around early SFGs \citep[e.g.,][]{Steidel2011,Hayes2013, Momose2014, Wisotzki2016, Wisotzki2018}, and there is a growing body of observational and computational work suggesting that \lya\ visibility in the early universe reflects, and is modulated by, the galactic environment on small and large scales \citep{Ouchi2010, Ouchi2018, Ouchi2020, Cooke2010, Cooke2013, Jose2013, Diaz2014, Muldrew2015, Toshikawa2016, Bielby2016, Guaita2017, Guaita2020, Lemaux2018, Shi2019}.  Given this established sensitivity of \lya\ to a wide range of intrinsic and environmental galactic properties, and the trends that have been demonstrated linking many of the same properties to galaxy kinematics, it is reasonable to ask the question: ``Is there a relationship between \lya\ and galaxy kinematics and how might this be used to inform our understanding of galaxy formation and evolution?" -- especially on large scales and at high redshifts where \lya\ is frequently the only spectroscopic indicator available. 

Radiative transfer simulations have investigated the influence of solid body rotation on \lya\ observables \citep{Garavito-Camargo2014, Remolina2019} and predict the sensitivity of \lya\ spectral line morphology to the bulk rotation of neutral gas and the viewing angle relative to the rotation axis.  They also predict, however, that there should be no observable difference in the integrated \lya\ line flux, the \lya\ escape fraction, or the average number of scatterings for each \lya\ photon caused by changes in the radiative transfer mechanism under the influence of rotation or dispersion-dominated kinematics alone.  The only direct observational study of a relationship between \lya\ and galaxy kinematics reported to date is the low-$z$ work of \citet{Herenz2016} who derived values for shear velocity and intrinsic velocity dispersion from the \halpha\ kinematic maps of galaxies in the \lymana\ Reference Sample \citep[LARS;][]{Ostlin2014, Hayes2014}.  The LARS collaboration surmise a causal connection between turbulence in actively star forming systems and interstellar medium conditions that favour an escape of \lya\ radiation, and further speculate that dispersion-dominated kinematics are a necessary requirement for a galaxy to have a significant amount of escaping \lya.

In the first paper in this series \citep[][hereafter referred to as Paper\,I]{Foran2023}, we report the photometric segregation of $z\sim2$ LBGs versus net \lya\ EW in colour-magnitude space, and derive criteria for the selection of pure samples of LBGs with \lya\ dominant in absorption and \lya\ dominant in emission using broadband imaging alone.  Together with the analogous $z\sim3$ result of \citet[][hereafter C09]{Cooke2009}, we have suggested the utility of this method to study a wide range of properties known to be associated with \lya\ in large samples and over large scales in data from current and future large-area photometric campaigns.  In particular, we foresee application of this approach to datasets from the all-sky LSST survey by the Vera C.\ Rubin Observatory that will select hundreds of millions of LBGs in redshift ranges from $z\sim2-6$ across many hundreds to thousands of Mpc.

In this paper we report a direct relationship between nebular emission-line kinematics and net \lya\ EW in  samples of $z\sim2$ and $z\sim3$ LBGs drawn from the literature, and extend the results of Paper\,I and C09 to propose a method by which the generalised kinematics of large samples of LBGs might be predicted using broadband imaging in as few as three filters, and studied on large scales in data from large-area and all-sky photometric surveys.  Finally, we combine our result with known relationships between \lya\ and galactic environment, and speculate on how these findings might be used to inform our understanding of galaxy formation and evolution in the early Universe.

This paper is structured as follows: In Section 2, we describe the photometric, spectroscopic, and kinematic data used in the subsequent sections.  A relationship between net \lya\ EW and the nebular emission-line kinematics of LBGs at $z\sim3$ and $z\sim2$ is presented in Section 3.  In Section 4, we discuss these results, their potential utility, and their implications for galaxy evolution science.  The important conclusions of the paper are summarised in Section 5.  We assume a $\Lambda$CDM cosmology with $\Omega_{M}$= 0.3, $\Omega_{\Lambda}$= 0.7 and H$_{0}$= 70\,km\,s$^{-1}$\,Mpc$^{-1}$.  All magnitudes are quoted in the AB system of \citet{Oke1983}.

\section{DATA} 
\label{sec:c4_data} 

\subsection{Overview}
\label{sec:c4_overview}

We assemble from the literature complementary kinematic samples of $z\sim2$ and $z\sim3$ LBGs with consistent multi-band rest-frame UV broadband photometry, uniformly measured net \lya\ EWs, and kinematic classifications quantitatively and comparably determined from IFU-based spectroscopy.  The $z\sim2$ kinematic sample consists of 23 UV-colour-selected (BX) LBGs and 13 $K_s$-band (mass) selected SFGs in the range $2.0 < z \lesssim 2.5$.  Twenty-two of these are classified as `rotation-dominated' (including all the $K_s$-band selected SFGs), four are `dispersion-dominated', and ten are classified as `mergers' in the source studies.  (see Section~\ref{sec:c4_z2kin} and Table~\ref{tab:c4_table2} for details).  The $z\sim3$ kinematic sample consists of 24 LBGs in the range $2.6 < z \lesssim 3.4$, of which  ten are classified as `rotating' or `rotation-dominated', eleven are `not-rotating' or `dispersion-dominated', and three galaxies are labelled as `not classified' in the source study (see Section~\ref{sec:c4_z3kin} and Table~\ref{tab:c4_table1} for details).  The LBGs in the kinematic samples all had broadband optical photometry available from the catalogs of \citet[][$z\sim2$]{Steidel2004} and \citet[][$z\sim3$]{Steidel2003}.  Broadband photometry for the $K_s$-band selected SFGs was transformed to match the $U_nG\cal{R}$ photometric system of these catalogs (see Section~\ref{sec:c4_kmos}).

The parent photometric catalogs derive from an observational campaign that targeted 14 uncorrelated fields with a total survey area of 1900 arcmin$^{2}$, resulting in a sample that is minimally affected by systematic biases due to cosmic variance or clustering.  The survey used the rest-frame UV colour selection criteria of \citet[][$z\sim3$]{Steidel2003} and \citet[][$z\sim2$]{Steidel2004, Adelberger2004}.  These criteria were designed to recover  galaxies with intrinsic properties -- particularly rest-frame UV luminosity and reddening by dust -- that were similar across both redshift ranges.  The faint end magnitude cuts of $\cal{R}$ $\leq 25.5$ (and $\cal{R}$ $\leq 26.0$ for one $z\sim3$ field) in the parent LBG samples were chosen so as to facilitate spectroscopic redshift determinations using the rich complement of strong interstellar and stellar  lines in the rest-frame UV continuum between \lya\ and $\sim$1700\,\AA\ \citep{Steidel2003, Steidel2004}.  Moreover, the faintest galaxies in the kinematic samples are brighter than $\cal{R}$ $=25$, thus mitigating any potential bias at the faint end due to over-reliance on \lya\ in emission for redshift determination.

The $z\sim2$ and $z\sim3$ parent samples have $\cal{R}$-band apparent magnitudes in the range $22.0 < \cal{R}$$ < 25.5$ and $22.7 < \cal{R} $$< 26.0$, corresponding to rest-frame UV luminosities (absolute magnitudes) of $-22.6 < \mathrm{M}_{UV} < -19.1$ and $-22.6 < \mathrm{M}_{UV} < -19.5$, respectively.  The bulk of galaxies in the parent samples have stellar masses in the range $9 \lesssim \mathrm{log}({M}_{\star }/{M}_{\odot }) \lesssim 11$ \citep{Shapley2003, Shapley2005, Erb2006b, Reddy2006, Reddy2009} and star formation rates (inferred from rest-frame UV luminosities uncorrected for extinction) in the range $3 \lesssim \mathrm{M}_{\odot}$\,yr$^{-1} \lesssim 60$ (median 9.9 $\mathrm{M}_{\odot}$\,yr$^{-1}$) and $5.5 \lesssim \mathrm{M}_{\odot}$\,yr$^{-1} \lesssim 66$, (median 10.3 $\mathrm{M}_{\odot}$\,yr$^{-1}$), respectively \citep{Steidel2004}.  Accordingly, our parent and kinematic samples are typical of LBGs/SFGs at these redshifts \citep[][and references therein]{Alvarez2016}, and the $z\sim2$ LBGs lie (though with a range of properties \citep[see][]{Reddy2006}) on the main sequence of stellar mass and star formation rate for $z\sim2$ SFGs \citep{Daddi2007}.

Throughout this work we use consistently determined net \lya\ EW as a measure of \lya\ visibility.  Net \lya\ EW incorporates information about \lya\ in emission, \lya\ in absorption, (even for strong to weak emitters) and their combined effects in the observed \lya\ spectral feature.  This is critically important for our LBG samples -- especially at $z\sim2$ where \lya\ in absorption dominates the population (see Paper\,I).  Net \lya\ EWs for galaxies in the parent LBG catalogs were measured uniformly at $z\sim2$ (see \citet{Reddy2008}) and $z\sim3$ \citep[][priv.\ comm.]{Shapley2003} using the method described by \citet{Kornei2010}.  The rest-frame UV colour criteria used to select the $z\sim2$ and $z\sim3$ LBGs result in a net \lya\ EW distribution for the $\cal{R}$ $<$ 25.5 samples that is representative of the intrinsic distribution for the parent population of galaxies \citep{Reddy2008}.

\subsection{$z\sim2$ Kinematic sample}
\label{sec:c4_z2kin} 

\subsubsection{Rest-frame UV-colour-selected galaxies}
\label{sec:c4_z2uv}

A sample of 23 rest-frame UV-colour-selected (BX) galaxies in the redshift range $2.0<z<2.5$ that overlap with our parent $z\sim2$ photometric catalog were selected from the SINS survey sample of \citet[][hereafter FS09]{FS2009} and the AO-assisted IFS survey of \citet[][hereafter LA09]{Law2009}.  Twenty-one of these galaxies had net \lya\ EWs in the spectroscopic catalog of \citet{Reddy2008}.

The BX galaxies targeted by FS09 and LA09 have stellar masses in the range $9.0 < \mathrm{log}({M}_{\star }/{M}_{\odot }) < 10.7$, and are drawn from the near-IR spectroscopic sample of \citet{Erb2006a, Erb2006c, Erb2006b}.  Although they have a number of galaxies in common, the LA09 galaxies tend to have stellar masses in the less-massive to typical-mass range (mean $\mathrm{log}({M}_{\star }/{M}_{\odot }) \approx 10.1$) compared to the FS09 sources that favour the higher mass end of the BX sample (mean $\mathrm{log}({M}_{\star }/{M}_{\odot }) \approx 10.42$). 

The SINS survey used the SINFONI instrument at the ESO VLT in natural-seeing and AO-assisted modes to extract spatially-resolved maps of the velocity-integrated flux, relative velocity, and velocity dispersion of the \halpha\ emission line.  To facilitate a general analysis of all the SINS \halpha\ galaxies, FS09 defined a working criterion involving the observed velocity gradient (${v}_{\mathrm{obs}}$) and the integrated line width (${\sigma}_{\mathrm{int}}$) by which galaxies with \vobs $> 0.4$ were classified as `rotation-dominated', and those with \vobs $< 0.4$ were classified as `dispersion-dominated'.  Using either quantitative kinemetric analysis \citep{Shapiro2008} or qualitative assessment of the asymmetry in the velocity field and dispersion map, galaxies with kinematics consistent with rotation were further classified as either `discs' or `mergers'.  Updated kinematic classifications for eight of the SINS objects were derived from the deep AO-assisted data collected as part of the SINS/zC-SINF survey \citep[][hereafter FS18]{FS2018}.  Galaxies identified by FS18 as possibly hosting an AGN (e.g, Q2343-BX610) were rejected from our sample.  

The LA09 AO study utilised the OSIRIS near-infrared integral field spectrograph \citep{Larkin2006} at the W.~M.~Keck Observatory.  LA09 quantified the rotational dynamic support using the ratio of shear velocity to intrinsic velocity dispersion (\vshear) and used detailed morphological analysis in combination with the 2D kinematic maps to characterise the kinematic properties of each galaxy.  For the purposes of kinematic classification, we equate \vshear\ with \vobs\ in the nomenclature of FS09, and, except for sources identified by LA09 as merging systems, assign the LA09 galaxies as either rotation or dispersion-dominated.

The grand design spiral galaxy Q2343-BX442 reported by \citet[][hereafter LA12]{Law2012b} was also included in our $z\sim2$ kinematic sample.  Net \lya\ EW and kinematic parameters for Q2343-BX442 were supplied by D.\ Law (priv.\ comm.).  With \vobs\ = 0.83, and clear disc-like morphology, we classify Q2343-BX442 as `rotation-dominated'.  Details of the $z\sim2$ UV-selected BX galaxies that comprise our kinematic sample are summarised in Table~\ref{tab:c4_table2} along with references to the source studies in each case.

\begin{table*}
\centering
\caption{$z\sim2$ SFGs used to establish the relationship between galaxy kinematics and net \lya\ EW in colour-magnitude space}
\label{tab:c4_table2}
\scalebox{0.77}{%
\begin{threeparttable}
\begin{tabular}{lccccccccc}
\toprule
\thead{ID} & 
\thead{RA} & 
\thead{DEC} & 
\thead{$z$} & 
\thead{$\cal{R}$-band\tnote{a} \\ Mag.} &
\thead{$(U_n-\cal{R})$\tnote{a} \\ Colour} & 
\thead{\lya\tnote{b} \\ type} & 
\thead{net \lya\\ EW (\AA)} &  
\thead{Kin.\ \tnote{c} \\ class.} & 
\thead{Ref.\tnote{d}} \\
\midrule
Q1623-BX376 & 16:25:45.59 & 26:46:49.26 & 2.4085 & 23.31 & 0.99 & aLBG & -12.03 & rd & FS09 \\
Q1623-BX447 & 16:25:50.37 & 26:47:14.28 & 2.1481 & 24.48 & 1.31 & aLBG & -10.05 & rd & FS09 \\
Q1623-BX455 & 16:25:51.66 & 26:46:54.88 & 2.4074 & 24.80 & 1.22 & G$_a$ & -6.54 & rd & FS09,NE13,FS18 \\
Q1623-BX502 & 16:25:54.38 & 26:44:09.25 & 2.1558 & 24.35 & 0.72 & eLBG & 28.49 & dd/rd\tnote{e} & FS09,LA09,FS18 \\
Q1623-BX528 & 16:25:56.44 & 26:50:15.44 & 2.2682 & 23.56 & 0.96 & aLBG & -11.53 & m & FS09 \\
Q1623-BX599 & 16:26:02.54 & 26:45:31.90 & 2.3304 & 23.44 & 1.02 & G$_a$ & -0.66 & m & FS09,NE13,FS18 \\
Q2343-BX389 & 23:46:28.90 & 12:47:33.55 & 2.1716 & 24.85 & 1.54 & G$_a$ & -8.30 & rd & FS09,NE13,FS18 \\
Q2343-BX513 & 23:46:11.13 & 12:48:32.14 & 2.1092 & 23.93 & 0.61 & G$_e$ & 18.43 & m & FS09,LA09,NE13,FS18 \\
Q2346-BX404\tnote{f} & 23:48:21.40 & 00:24:43.07 & 2.0282 & 23.39 & 0.59 & G$_e$ & 2.39 & dd & FS09 \\
Q2346-BX405\tnote{f} & 23:48:21.22 & 00:24:45.46 & 2.0300 & 23.36 & 0.68 & G$_a$ & -8.61 & dd & FS09 \\
Q2346-BX416 & 23:48:18.21 & 00:24:55.30 & 2.2404 & 23.49 & 1.15 & aLBG & -17.22 & rd & FS09 \\
Q2346-BX482 & 23:48:12.97 & 00:25:46.34 & 2.2569 & 23.32 & 1.12 & aLBG & -14.98 & rd & FS09,NE13,FS18 \\
Q0449-BX93 & 04:52:15.417 & -16:40:56.88 & 2.0067 & 22.99 & 0.63 & aLBG & -13.31 & m & LA09 \\
Q1217-BX95 & 12:19:28.281 & 49:41:25.90 & 2.4244 & 24.11 & 1.21 & G$_e$ & 10.17 & dd & LA09 \\
HDF-BX1564 & 12:37:23.47 & 62:17:20.02 & 2.2225 & 23.28 & 1.28 & aLBG & -11.12 & m & LA09 \\
Q1623-BX453 & 16:25:50.84 & 26:49:31.40 & 2.1816 & 23.38 & 1.47 & G$_a$ & -0.16 & m & LA09 \\
Q1700-BX490 & 17:01:14.83 & 64:09:51.69 & 2.3960 & 22.88 & 1.28 & G$_e$ & 5.90 & m & LA09 \\
Q2343-BX418 & 23:46:18.57 & 12:47:47.38 & 2.3052 & 23.99 & 0.32 & eLBG & 53.49 & dd & LA09 \\
Q2343-BX660 & 23:46:29.43 & 12:49:45.54 & 2.1735 & 24.36 & 0.36 & eLBG & 20.38 & m & LA09 \\
Q1623-BX543 & 16:25:57.70 & 26:50:08.59 & 2.5211 & 23.11 & 1.40 & G$_a$ & -7.80 & m & FS09,LA09,NE13,FS18 \\
Q1623-BX663 & 16:26:04.58 & 26:48:00.20 & 2.4333 & 24.14 & 1.26 & -- & -- & m & FS09,NE13 \\
SSA22a-MD41 & 22:17:39.97 & 00:17:11.04 & 2.1713 & 23.31 & 1.50 & -- & -- & rd & FS09 \\
Q2343-BX442 & 23:46:19.36 & 12:47:59.69 & 2.1760 & 24.48 & 1.54 & aLBG & -15.0 & rd & LA12 \\
COS4\_04519 & 10:00:28.65 & 02:13:27.0 & 2.2290 & 23.80$^*$ & 1.46$^*$ & G$_a$ & -9.1 & rd & WI15 \\
COS4\_05094 & 10:00:33.69 & 02:13:48.5 & 2.1720 & 24.72$^*$ & 1.47$^*$ & -- & -- & rd & WI15 \\
COS4\_05389\tnote{g} & 10:00:17.59 & 02:13:58.7 & 2.3060 & 23.91$^*$ & 1.26$^*$ & -- & -- & rd & WI15 \\
COS4\_08775 & 10:00:16.54 & 02:16:09.3 & 2.1630 & 23.42$^*$ & 0.95$^*$ & aLBG & -13.2 & rd & WI15 \\
COS4\_10347 & 10:00:20.03 & 02:17:08.8 & 2.0630 & 23.85$^*$ & 1.44$^*$ & -- & -- & rd & WI15 \\
COS4\_13174 & 10:00:26.96 & 02:18:50.0 & 2.0980 & 26.20$^*$ & 1.72$^*$ & -- & -- & rd & WI15 \\
COS4\_13701 & 10:00:27.05 & 02:19:09.9 & 2.1670 & 23.62$^*$ & 1.50$^*$ & aLBG/G$_a$ & <0.0 & rd & WI15 \\
COS4\_15813 & 10:00:34.98 & 02:20:28.8 & 2.3590 & 24.14$^*$ & 1.87$^*$ & aLBG & -15.1 & rd & WI15 \\
COS4\_15820 & 10:00:15.56 & 02:20:29.7 & 2.0910 & 23.63$^*$ & 1.16$^*$ & -- & -- & rd & WI15 \\
COS4\_16342 & 10:00:29.41 & 02:20:49.6 & 2.4690 & 24.27$^*$ & 2.09$^*$ & aLBG/G$_a$ & <0.0 & rd & WI15 \\
COS4\_18209 & 10:00:20.16 & 02:21:53.3 & 2.0870 & 23.74$^*$ & 2.14$^*$ & aLBG/G$_a$ & <0.0 & rd & WI15 \\
COS4\_19680 & 10:00:18.06 & 02:22:45.8 & 2.1670 & 24.54$^*$ & 1.37$^*$ & aLBG/G$_a$ & <0.0 & rd & WI15 \\
COS4\_24763 & 10:00:13.61 & 02:26:04.8 & 2.4650 & 24.08$^*$ & 1.97$^*$ & -- & -- & rd & WI15 \\
\bottomrule
\end{tabular}
\begin{tablenotes}
\item [a] Magnitudes and colours marked with an asterisk ($^*$) are derived from \ugriz\ photometry using the method outlined in Section~\ref{sec:c4_kmos}.
\item [b]  See Section~\ref{sec:c4_lya_type} for definitions of \lya\ spectral types.
\item [c]  Kinematic classification: rd = `rotation-dominated', dd = `dispersion-dominated' and m = `merger' according to the criteria described in Section~\ref{sec:c4_z2kin}.
\item [d]  Source references for kinematic classifications: FS09 = \citet{FS2009}, LA09 = \citet{Law2009}, LA12 = \citet{Law2012b}, NE13 = \citet{Newman2013}, WI15 = \citet{Wisnioski2015}, FS18 = \citet{FS2018}
\item[e]  See Section~\ref{sec:c4_elbgs} for notes regarding the kinematic classification of Q1623-BX502.
\item[f]  Q2346-BX404 and Q2346-BX405 are an interacting pair with angular separation corresponding to a projected distance of 30.3 kpc at the redshift of the sources (FS09).  
\item[g]  Photometry for COS4\_05389 is derived from that of the two CFHTLS sources of which it is comprised (CFHTLS IDs 259119 and 259367).
\end{tablenotes}
\end{threeparttable}}
\end{table*}

\subsubsection{KMOS$^{\mathrm{3D}}$ Galaxies} 
\label{sec:c4_kmos}

We supplement our $z\sim2$ kinematic sample with a redshift-selected subset ($2.0<z<2.5$) of 13 galaxies from the COSMOS field pointings of KMOS$^{\mathrm{3D}}$ (\KMOS) -- an integral field survey of over 600 mass-selected galaxies at 0.7 $<z<$ 2.7 using the KMOS instrument at the ESO VLT \citep{Wisnioski2019}.  The \KMOS\ survey combined galaxy dynamics derived from \halpha, near-IR continuum, velocity, and velocity dispersion maps with structural parameters and multi-band imaging to establish a set of criteria by which robust kinematic classifications could be determined \citep{Wisnioski2015}.  The \KMOS\ sources that we employ are part of the `high S/N disc sample' of rotation-dominated galaxies reported by \citet[][hereafter WI15]{Wisnioski2015} that focused on massive galaxies with $\mathrm{log}({M}_{\star }/{M}_{\odot }) \gtrsim 10$  (see Table~\ref{tab:c4_table2} for details).  These galaxies all meet the less exacting FS09 criterion of \vobs\ > 0.4 for classification as rotation-dominated in our study.  The \KMOS galaxies were cross-matched with the D2 field of the Canada-France-Hawaii Telescope Legacy Survey \citep[CFHTLS][]{Hudelot2012} to obtain $u^* g^\prime r^\prime i^\prime z^\prime$ multi-band photometry.  The \ugriz\ photometric data were transformed into \UGR\ magnitudes to facilitate direct comparison with the $z\sim2$ sources of FS09, LA09, FS18 and LA12.

The transformation was achieved by performing spectrophotometry on rest-frame UV composite spectra derived from a sample of $z\sim3$ \UGR\ LBGs divided into quartiles on the basis of net \lya\ EW (A.\ Shapley, priv.\ comm.).  The composite spectra are representative of the average LBG spectrum in each quartile, and thus accurately trace the colour--colour evolution and colour--magnitude distribution of each quartile, and its component galaxies \citep{Cooke2013}.  For each \KMOS\ galaxy, the composite spectra were first redshifted to the observation frame, and flux density in the \lymana\ forest corrected for redshift-dependent absorption through the IGM.  The spectra were then convolved with the bandpasses of the $u^* g^\prime r^\prime$ filters, the resulting integrated flux densities normalised to the observed $g^\prime $-band magnitude, and simulated $u^* g^\prime r^\prime$-band magnitudes calculated.  From these, the composite spectrum that best fit the observed \ugriz\ photometry for each galaxy was determined.  A reddening correction \citep{Calzetti2000} was applied to the best fit spectrum in each case as required to optimise the fit.  The best fit normalised and reddened spectrum was then convolved with the bandpassess of the \UGR\ filters, and \UGR-band magnitudes estimated for each galaxy.  In all cases, the quartile 1 (strongest net \lya\ EW absorption) or the quartile 2 (next strongest net \lya\ EW absorption with some emission) composite spectra provided the best fit to the observed photometry, suggesting that the \KMOS\ galaxies are best classified as aLBG or G$_a$ spectral types (see Section~\ref{sec:c4_lya_type}).  Using the transformed \UGR\ photometry, a standard colour--colour test \citep{Steidel2004, Adelberger2004} was applied to confirm that the \KMOS\ sample satisfied (within the photometric uncertainties) the criteria to be selected as $z\sim2$ LBGs.

With the goal of measuring net \lya\ EWs, seven of the `high S/N disc sample' \KMOS\ galaxies were included as secondary science targets on our multi-object slitmasks using the LRIS instrument \citep{Oke1995,Steidel2004} at Keck on 26, 27 December 2016 and 20--22 January 2020.  These data were reduced in the conventional manner using IRAF and in-house code.  Net \lya\ EWs were measured following the procedure of \citet{Kornei2010}.  We successfully measured net \lya\ EW for three WI15 galaxies from the January 2020 data. However, weather conditions and primary science constraints during the December 2016 run resulted in low S/N spectra for the remaining four WI15 galaxies that were too poor to enable the reliable measurement of net \lya\ EW.  The 2D and 1D spectra of these four galaxies show no evidence of a \lya\ emission component of $\gtrsim$10~\AA.  The q1 and q2 quartiles of \citet{Shapley2003} have average net \lya\ EWs $-$14.9\,\AA\ and $-$1.1\,\AA, respectively, and show \lya\ emission components of $<$ 10\,\AA.  Consequently, we can reasonably postulate that the net \lya\ EWs for the four galaxies are similar to the q1 and q2 quartile galaxies.  On this basis, we report net \lya\ EW $<0.0$ for these four sources and provisionally assign them spectral types aLBG/G$_a$ in our system

\subsection{$z\sim3$ Kinematic sample}
\label{sec:c4_z3kin} 

\subsubsection{AMAZE and LSD Galaxies}
\label{sec:c4_amaze}

Using the SINFONI integral field spectrograph \citep{Eisenhauer2003} on the ESO Very Large Telescope (VLT) in natural-seeing and natural guide star adaptive optics (AO) observation modes respectively, the related AMAZE \citep{Maiolino2008} and LSD \citep{Mannucci2009} surveys conducted near-IR IFU spectroscopic observations on LBGs at redshifts $z\gtrsim3$ with $\cal{R}$ $\simeq$ 24.5 (L* and brighter) corresponding to a mass range of $\mathrm{log}({M}_{\star }/{M}_{\odot })\approx 10-11$.  

\citet[][hereafter GN11]{Gnerucci2011} derived nebular emission-line kinematics for a subset of amaze and lsd galaxies by fitting the profile and shift of the  [O{\footnotesize III}]$\lambda\lambda$4959,5007 doublet.  Due to limited signal-to-noise, GN11 used a plane-fitting method to assign kinematic classifications  according to the following criteria: galaxies for which the velocity map shows a non-zero gradient after plane-fitting were classified as `rotating'; galaxies for which the velocity map could not be fitted with a plane were classified as `not-rotating'; and galaxies with velocity maps well-fitted by a plane but with inclination consistent with zero were labelled as `not classifiable'.  GN11 further employed a rotating-disc modelling approach to estimate maximum rotation velocities and intrinsic velocity dispersions for galaxies classified as `rotating' in their sample. 

We extract 18 GN11 galaxies that overlap with our parent photometric catalog.  Details of the AMAZE and LSD galaxies used in this work are summarised in Table~\ref{tab:c4_table1}.

\subsubsection{KDS galaxies} 
\label{sec:c4_kDS}

As part of the KMOS Deep Survey (KDS), \citet[][hereafter TU17]{Turner2017} investigated the kinematics of typical isolated field SFGs at z $\simeq$ 3.5 in the mass range $9.0 < \mathrm{log}({M}_{\star }/{M}_{\odot }) < 10.5$ using the KMOS instrument at the ESO VLT \citep{Sharples2013}.  With natural-seeing measurements of the [O{\footnotesize III}]$\lambda$5007 emission line, TU17 extracted 2D kinematic maps and used beam-smearing corrections derived from dynamical modelling to determine values of intrinsic rotation velocity (V$_\mathrm{C}$) and intrinsic velocity dispersion (${\sigma}_\mathrm{int}$) for the spatially-resolved target galaxies in their sample.  Dictated by the signal-to-noise ratio of their data, TU17 used a simple empirical diagnostic based on the ratio V$_\mathrm{C}$/$\sigma_\mathrm{int}$ to kinematically classify their sample.  Galaxies were classified as `rotation-dominated' if V$_\mathrm{C}$/$\sigma_\mathrm{int} > 1$, and as `dispersion-dominated' if V$_\mathrm{C}$/$\sigma_\mathrm{int} < 1$. 

The SSA22-P2 pointing of the KDS survey targeted a field environment  to the south of the main SSA22 spatial overdensity \citep{Steidel2000} and yielded five morphologically isolated (non-merger) field galaxies that were in common with our parent photometric catalog.  Details of the five  KDS galaxies used in this work are given in Table~\ref{tab:c4_table1}.

\subsubsection{DSF2237a-C2}
\label{sec:c4_DSF2237a-C2}

With a redshift of $z \simeq 3.3$ and kinematics derived from measurements of the [O{\footnotesize III}]$\lambda$5007 emission line, we include the LA09 galaxy, DSF2237a-C2, in our $z\sim3$ kinematic sample.  Drawn from the $z\sim3$ LBG catalog of \citet{Steidel2003}, DSF2237a-C2 has $v_{\mathrm{shear}}/{\sigma}_{\mathrm{mean}} = 0.6 \pm 0.2$, and is the only isolated LA09 galaxy where the observed velocity gradient "is consistent with rotation and unambiguously aligned with the morphological major axis" (LA09).  For the purposes of kinematic classification, we treat DSF2237a-C2 similarly to the $z\sim2$ LA09 galaxies (see Section~\ref{sec:c4_z2uv}), and assign a `rotation-dominated' classification to this galaxy.

\begin{table*}
\centering
\caption{$z\sim3$ LBGs used to establish the relationship between galaxy kinematics and \lya\ EW in colour-magnitude space}
\label{tab:c4_table1}
\scalebox{0.8}{%
\begin{threeparttable}
\begin{tabular}{lcccccccccc}
\toprule
\thead{ID} & 
\thead{RA} & 
\thead{DEC} & 
\thead{$z$} & 
\thead{$\cal{R}$ Mag.} &
\thead{$U_n - \cal{R}$\\ Colour} & 
\thead{\lya \\type \tnote{a}} & 
\thead{net \lya\\ EW (\AA)} &  
\thead{Kin.\ \tnote{b} \\ class.} & 
\thead{Obs.\tnote{c} \\ mode} & 
\thead{Ref.\tnote{d}} \\
\midrule
SSA22a-M38 & 22:17:17.69 & +00:19:0.70 & 3.2880 & 24.11 & 1.15 & aLBG & -16.52 & rot & NS & GN11 \\
SSA22a-C16 &    22:17:31.95 &   +00:13:16.10 &  3.0650 &   23.64 &    0.98 & aLBG &  -21.24 & rot & NS &  GN11 \\
SSA22a-D17 & 22:17:18.87 & +00:18:16.80 & 3.0890 & 24.27 & 0.45 & G$_e$  & 18.14 & rot & NS & GN11 \\
CDFA-C9 & 00:53:13.71 & +12:32:11.10 & 3.2095 & 23.99 & 0.89 & G$_a$  & -6.48 & rot & NS & GN11 \\
3C324-C3 &    15:49:47.10 &    +21:27:5.40 &  3.2890 &   24.14 &    0.85 &   G$_e$  & 0.97 & rot &   NS &  GN11 \\
DSF2237b-D29 & 22:39:32.70 & +11:55:51.70 & 3.3670 & 23.70 & 0.96 & G$_e$  & 1.79 & not-rot & NS & GN11 \\
DSF2237b-C21 & 22:39:29.00 & +11:50:58.10 & 3.4030 & 23.50 & 1.10 & G$_a$  & -0.12 & not-rot & NS & GN11 \\
SSA22a-aug96M16 & 22:17:30.85 & +00:13:10.70 & 3.2920 & 23.83 & 0.70 &  G$_e$ & 19.40 & not-rot & NS & GN11 \\
SSA22a-C36 & 22:17:46.09 & +00:16:43.00 & 3.0630 & 24.06 & 0.78 & aLBG & -13.67 & NC & NS & GN11 \\
SSA22a-D3 &    22:17:32.41 &     00:11:32.9 &  3.0860 &   23.37 &    0.97 &  G$_e$  & 11.47 & rd &   NS &  TU17 \\
SSA22b-C20 &    22:17:48.81 &     00:10:14.0 &  3.2020 &   24.60 &    0.68 &  G$_a$  &   -4.18 & rd &   NS &  TU17 \\
SSA22b-D5 &    22:17:35.75 &     00:06:10.5 &  3.1880 &   24.42 &    0.66 & G$_e$ &  1.04 & rd & NS & TU17 \\
SSA22b-D9 &    22:17:22.27 &     00:08:04.2 &  3.0880 &   25.14 &    0.21 & eLBG &   37.83 & dd &   NS &  TU17 \\
SSA22b-MD25 &    22:17:41.65 &     00:06:20.5 &  3.3090 &   24.53 &    0.56 & eLBG &   32.81 & dd &   NS &  TU17 \\
SSA22b-C5 &    22:17:47.06 &   +00:04:25.70 &  3.1140 &   25.16 &    0.71 & eLBG &   85.25 & not-rot & AO &  GN11 \\
SSA22a-C6 &    22:17:40.93 &   +00:11:26.00 &  3.0950 &   23.44 &    0.79 &  G$_e$  & 6.19 & not-rot\tnote{e} & AO &  GN11 \\
SSA22a-M4 &    22:17:40.91 &   +00:11:27.90 &  3.0930 &   24.83 &    0.76 &  G$_a$  &   -7.70 & NC\tnote{e} & AO &  GN11 \\
SSA22a-C30 &    22:17:19.28 &   +00:15:44.70 &  3.1010 &   24.22 &    0.82 &   G$_e$  &    3.60 & not-rot &   AO &  GN11 \\
Q0302-C131 & 03:04:35.04 &  -00:11:18.30 & 3.2340 &  24.48 & 0.54 & eLBG & 27.60 & rot & AO &  GN11 \\
Q0302-C171 &    03:04:44.31 &   -00:08:23.20 &  3.3350 &   24.63 &   0.60 &  G$_a$  &   -0.91 & NC &   AO &  GN11 \\
DSF2237b-D28 &    22:39:20.25 &   +11:55:11.30 &  2.9320 &   24.46 &    0.32 &   G$_e$  &    1.03 & not-rot &   AO &  GN11 \\
Q0302-M80 &    03:04:45.70 &   -00:13:40.60 &  3.4160 &   24.12 &    1.01 &   G$_e$  &    0.81 & not-rot &   AO &  GN11 \\
DSF2237b-MD19 &    22:39:21.08 &   +11:48:27.70 &  2.6150 &   24.48 &    0.44 &  G$_e$  &   13.94 & not-rot & AO &  GN11 \\
DSF2237a-C2 &     22:40:8.29 &    +11:49:4.80 &  3.3172 &   23.55 &    1.13 &  G$_e$  &    0.03 & rd & AO & LA09 \\
\bottomrule
    \end{tabular}
\begin{tablenotes} 
\item [a]  See Section~\ref{sec:c4_lya_type} for definitions of \lya\ spectral types.
\item [b]  Kinematic classification: rot = `rotating', not-rot = `not-rotating', rd = `rotation-dominated', dd = `dispersion-dominated' and NC = `not classifiable' according to the criteria of the respective source studies (see Section~\ref{sec:c4_z3kin}).
\item [c]  NS = natural seeing, AO = adaptive optics assisted
\item [d]  Source references for kinematic classifications: GN11 = \citet{Gnerucci2011}, LA09 = \citet{Law2009}, TU17 = \citet{Turner2017}
\item [e]  GN11 identifies SSA22a-C6 and SSA22a-M4 as a close-pair.  They have a projected separation of 12 kpc.  and SSA22a-M4 is redshifted by $\sim$90\kms\ relative to SSA22a-C6. 
\end{tablenotes}
\end{threeparttable}}
\end{table*}

\subsection{\lya\ spectral type classifications}
\label{sec:c4_lya_type}

We employ herein the same \lya\ spectral type classification scheme as that used by C09 at $z\sim3$ and in Paper\,I at $z\sim2$ to demonstrate the photometric segregation of LBGs with respect to net \lya\ EW.  For this purpose we define galaxies with \lya\ dominant in absorption (net \lya\ EW $<$ $-$10.0\AA) as `aLBGs', and galaxies with \lya\ dominant in emission (net \lya\ EW $>$ 20.0 \AA) as `eLBGs'.  We further divide the remaining $z\sim2$ LBGs into G$_a$ and G$_e$ spectral types with net \lya\ EWs in the range $-$10.0 $<$ net \lya\ EW $<$ 0.0\,\AA\ and 0.0 $<$ net \lya\ EW $<$ 20.0\,\AA\ respectively.  

\section{ANALYSIS AND RESULTS}

\subsection{$z\sim$3 LBGs}
\label{sec:c4_z3lbgs}

\begin{figure*}
\centering
\scalebox{0.58}[0.58]{\includegraphics{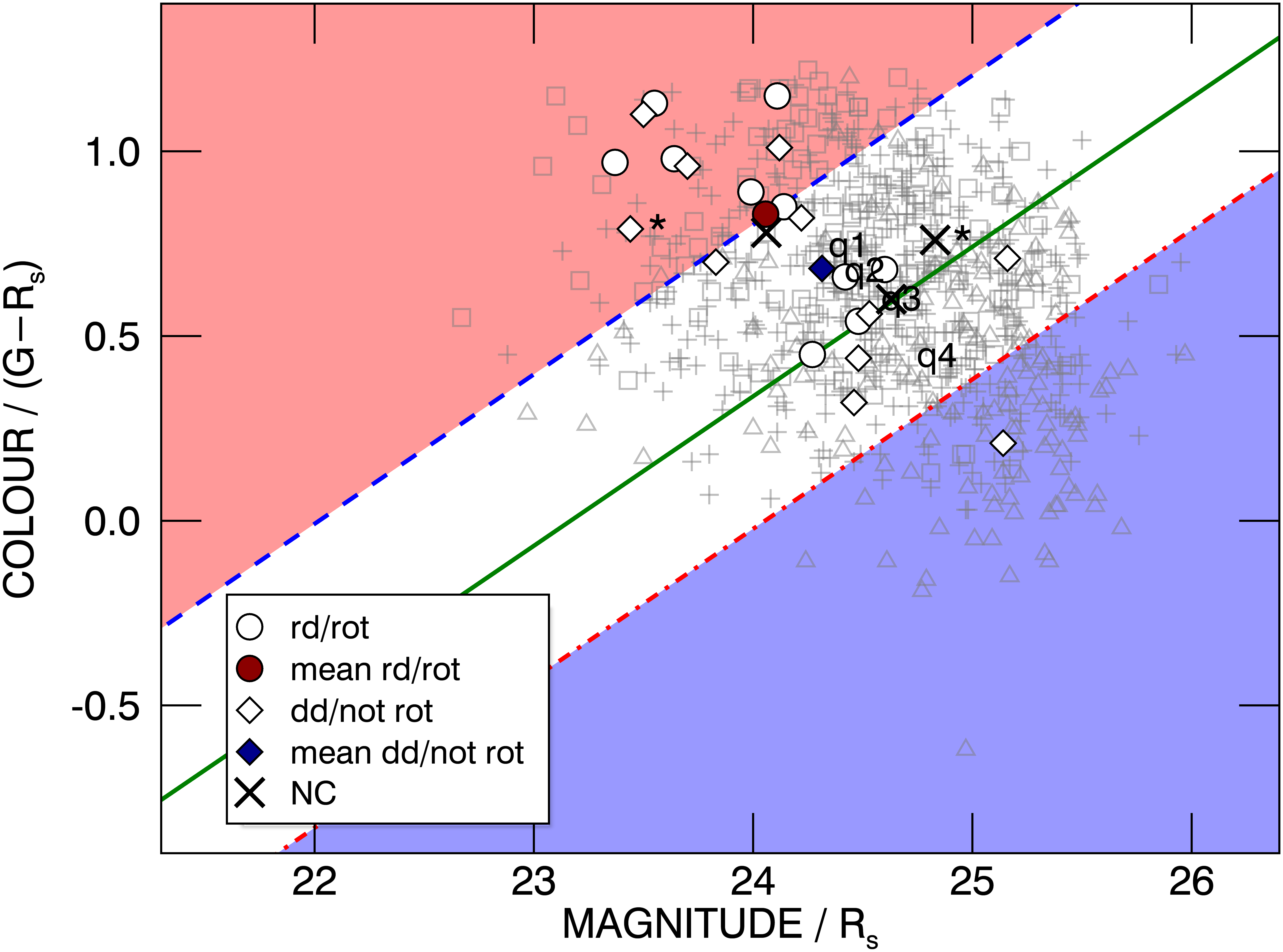}}
\scalebox{0.58}[0.58]{\includegraphics{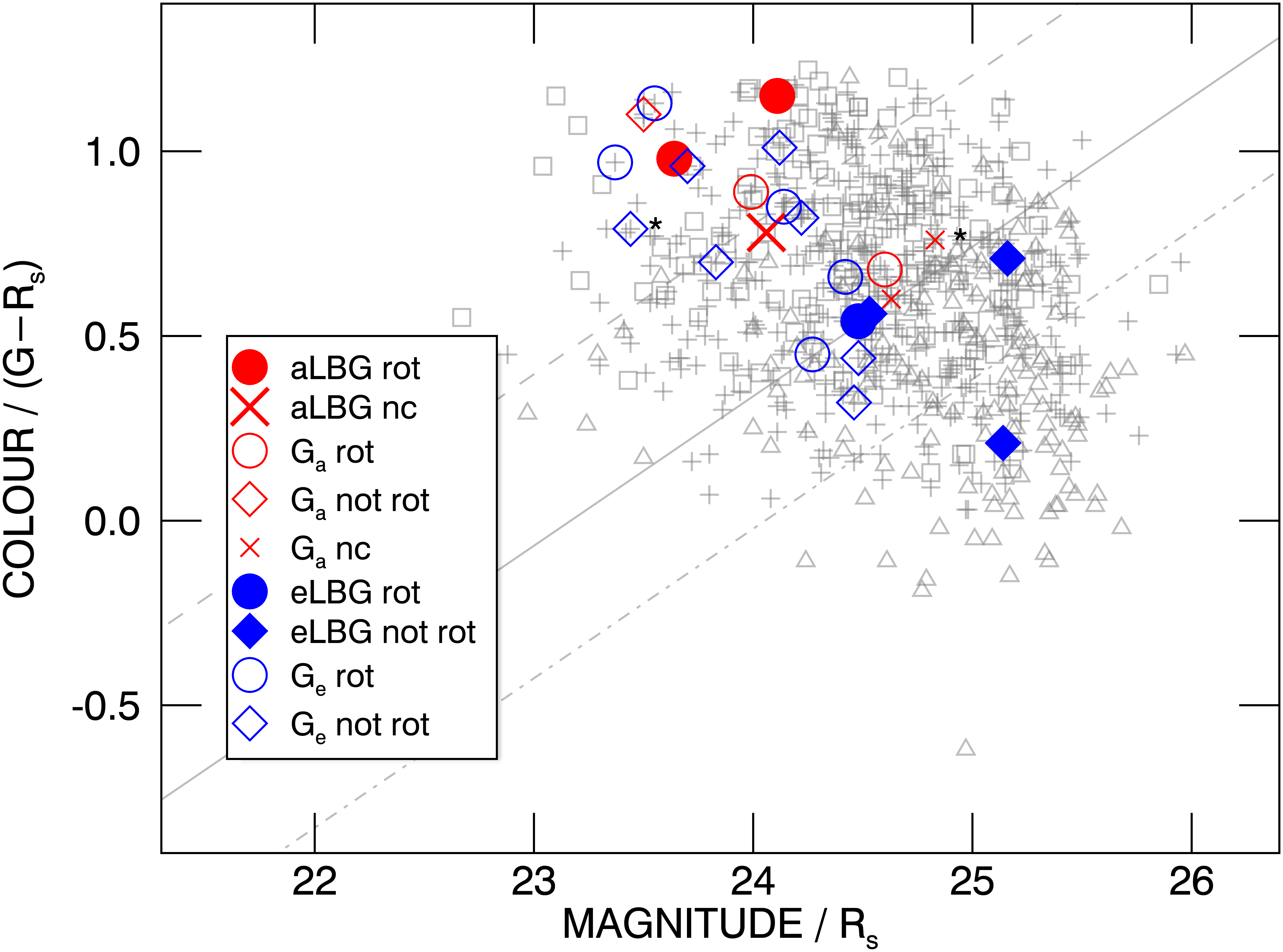}} 
\caption{Rest-frame UV colour--magnitude ($(G-\cal{R})$ vs $\cal{R}$) diagrams adapted from \citet{Cooke2009}. In both plots, the parent sample of 775 spectroscopic $z\sim3$ LBGs is shown in light grey: squares indicate aLBGs, triangles eLBGs, and plus signs galaxies with intermediate values of net \lya\ EW.  Left: Points labelled q1--q4 mark the colour and magnitude distribution means of the parent LBG sample divided into numerical quartiles on the basis of net \lya\ EW.  The primary cut (green line) bisects the aLBG and eLBG distributions.  The dashed (red) and dotted-dashed (blue) lines indicate an offset of 1.5$\sigma$ in colour dispersion from the primary cut for the aLBG and eLBG distributions respectively, and define the shaded red and blue regions within which pure samples of \lya-absorbing and \lya-emitting LBGs can be selected with high confidence (see text).  The $z\sim3$ kinematic sample is overlaid as white symbols.  Galaxies classified as `rotation-dominated' or `rotating' (rd/rot) are shown as circles, and galaxies classified as `dispersion-dominated' or `not-rotating' (dd/not rot) as diamonds.  The mean colour--magnitude values of the rotation-dominated/rotating and dispersion-dominated/not-rotating sub-samples are indicated by the red circle and blue diamond respectively.  Black crosses (X) are galaxies designated as `not classifiable' by GN11, and galaxies marked with an asterisk ($^*$) are members of an interacting close-pair.  Right: Similar to the left panel, but with \lya\ spectral types identified.  Filled red symbols denote aLBGs, filled blue symbols are eLBGs, and hollow red and blue symbols are G$_a$ and G$_e$ spectral types respectively (see Section~\ref{sec:c4_lya_type}). The distribution of the kinematic sample on the CMD reflects the selection bias toward Brighter (more massive) galaxies in the source IFU studies.}
\label{fig:c4_fig1}
\end{figure*}

C09 discovered that $z\sim3$ LBGs segregate in colour--magnitude space according to their net \lya\ EW, and determined photometric criteria to select pure sub-samples with \lya\ dominant in absorption (aLBGs) and \lya\ dominant in emission (eLBGs) based on broadband imaging.  Figure~\ref{fig:c4_fig1} shows our $z\sim3$ kinematic sample overlaid on rest-frame UV colour--magnitude diagrams (CMDs) of $z\sim3$ LBGs segregated according to their \lya\ spectral type adapted from C09.  In the left panel, points labelled q1-q4 show the monotonic trend of the parent sample divided into numerical quartiles on the basis of net \lya\ EW.  The more positive net \lya\ EW quartiles (weaker absorption and stronger emission) trend consistently toward fainter $\cal{R}$-band magnitudes and bluer $(G-\cal{R})$ colours.  The primary cut (solid green line) statistically divides the mean colour and magnitude values of the aLBG and eLBG distributions.  The dashed (red) and dotted-dashed (blue) lines indicate an offset of 1.5$\sigma$ in colour dispersion from the primary cut for the aLBG and eLBG distributions respectively, and define one choice of photometric criteria for the selection of pure \lya-absorbing and \lya-emitting sub-samples that lie in the shaded red and blue regions  respectively.  

The aLBG and eLBG distributions segregate on the CMD such that $\gtrsim90$~percent of aLBGs (grey squares) are located above the red dotted-dashed line (in the white and red regions), and $\lesssim10$~percent of eLBGs (grey triangles) are located in the red region above the blue dashed line.  The reverse is true for eLBGs; $\gtrsim90$~percent of eLBGs lie below the dashed line (in the white and blue regions), and $\lesssim10$~percent of aLBGs are found in the blue region below the dotted-dashed line.  Thus, the aLBG and eLBG distributions partially overlap in the central part of the CMD, but relatively pure subsets of aLBGs and eLBGs can be selected from the `high-confidence' red and blue regions respectively (see C09 and Paper\,I for quantitative details of the \lya\ spectral type photometric selection method).

Galaxies in the $z\sim3$ kinematic sample classified as `rotating' or `rotation-dominated' lie at or above the primary cut coincident with the majority of aLBGs.  Galaxies classified as `not-rotating' or `dispersion-dominated' are scattered on the CMD, but all galaxies below the primary cut, and in the `high-confidence' eLBG (blue) region are classified as `not-rotating' or `dispersion-dominated'.  In addition, the not-rotating/dispersion-dominated sub-sample is on average fainter and bluer than their rotating/rotation-dominated counterparts, as indicated by the red circle and blue diamond in the left panel of Figure~\ref{fig:c4_fig1}, and we note that the `not-rotating' sub-sample may include late-stage mergers (GN11).  Thus, the rotating/rotation-dominated and not-rotating/dispersion-dominated subsets of the kinematic sample follow the aLBG and eLBG distributions within their known dispersion characteristics.

This trend is reinforced when we include the spectroscopically-determined net \lya\ EW data and assign a \lya\ spectral type to each galaxy in our kinematic sample according to the definitions given in Section~\ref{sec:c4_lya_type}.  The right panel of Figure~\ref{fig:c4_fig1} shows the $z\sim3$ kinematic sample colour-coded according to their \lya\ spectral type overlaid on the parent LBGs.  The three aLBGs lie well above the primary cut; of these, two are confirmed rotators, and the third is `not classifiable' and could be a face-on disc (GN11).  The four eLBGs, of which three are `not-rotating' or `dispersion-dominated', lie on or below the primary cut.  The association between kinematics and \lya\ spectral type is less clear for intermediate G$_a$ and G$_e$ LBGs -- particularly for those that lie toward the centre of the CMD -- and may be obscured by the `rotating'/`not-rotating' classification scheme of the source study that precludes an interpretation in terms of late-stage merging systems (GN11).

Of the ten not-rotating/dispersion-dominated galaxies in the $z\sim3$ kinematic sample, three are eLBGs, six have G$_e$ spectral type, and one is borderline G$_a$ with a net \lya\ EW of $-$0.1~\AA.  The mean net \lya\ EW for these 10 galaxies is $+$19.6~\AA.  The ten rotating/rotation-dominated sources consist of two aLBGs, two G$_a$ LBGs, three borderline G$_a$/G$_e$ spectral types with net \lya\ EW $\approx 0$~\AA, two G$_e$ LBGs, and one eLBG.  Two `rotating' galaxies (SSA22a-D17 and Q0302-C131) have significant net \lya\ emission ($+$18.14 and $+$27.60~\AA\ respectively).  The average net \lya\ EW for the rotating/rotation-dominated sub-sample is $+$1.1~\AA.

These results suggest an empirical relationship between rest-frame UV colour, \lya\ spectral type, and the kinematic properties of $z\sim3$ LBGs.  In the following section we extend our study to $z\sim2$ where a larger number of LBGs with kinematic classifications, appropriate imaging, and \lya\ spectroscopic data are available.

\subsection{$z\sim$2 LBGs}
\label{sec:c4_z2lbgs}

\subsubsection{$z\sim2$ Kinematics on the CMD}
\label{sec:c4_z2kincmd}

\begin{figure}
\centering
\includegraphics[width=\columnwidth]{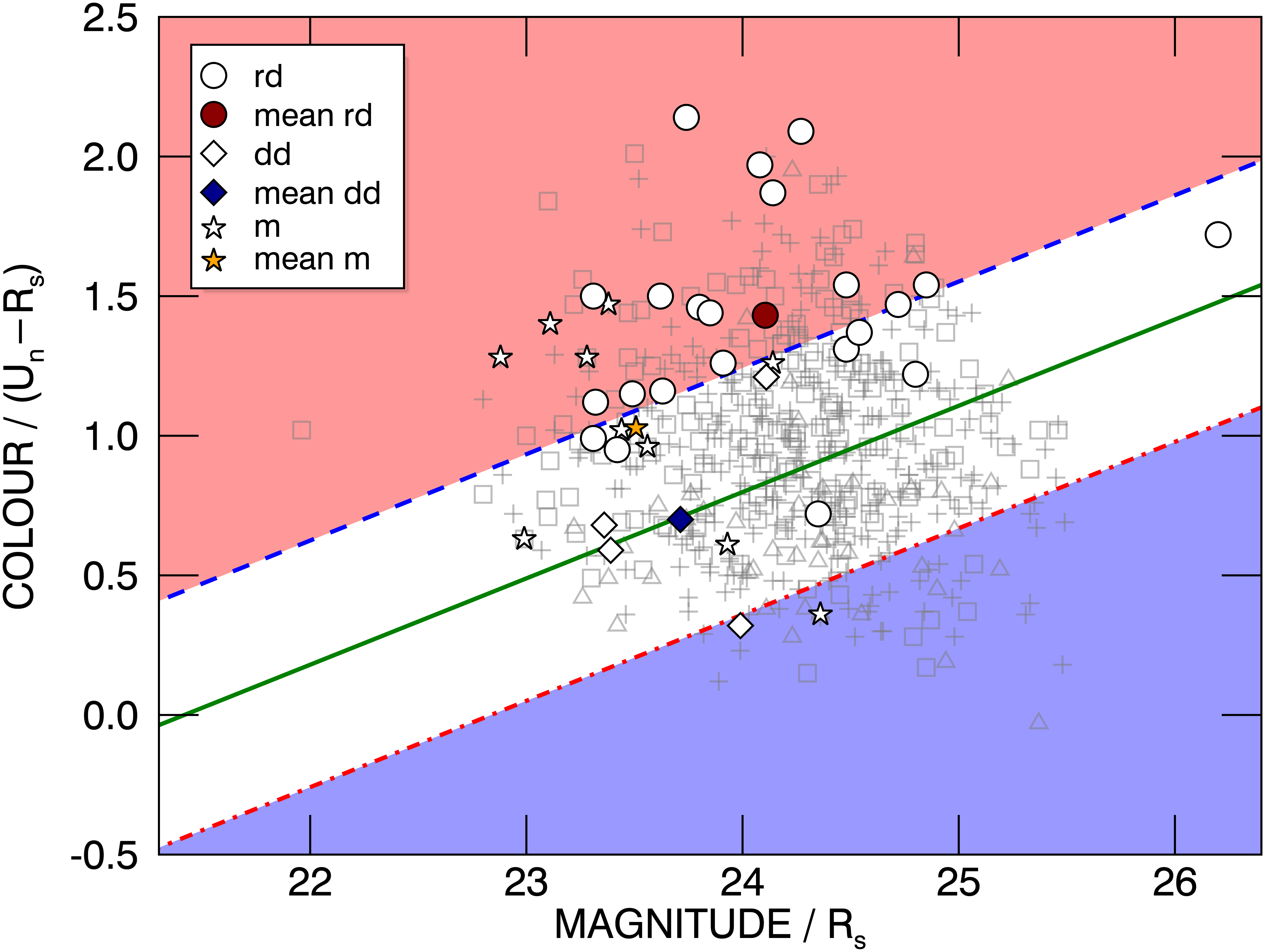}
\caption{Similar to the left panel of Fig.~\ref{fig:c4_fig1} but for the $z\sim2$ kinematic sample plotted on a $(U_n-\cal{R})$ vs.\ $\cal{R}$ CMD adapted from Paper\,I.  Galaxies classified as `rotation-dominated' (rd, circles) follow the form of the aLBG population distribution, and galaxies classified as `dispersion-dominated' (dd, diamonds) follow the eLBG population to the extent that can be estimated given the scarcity of \lya-emitting galaxies in the sample, and at $z\sim2$ in general.  Galaxy mergers (m, stars) are typically found central in colour and toward the bright end of the CMD.  The mean positions of the  rd, dd, and m sub-samples are marked by a red circle, blue diamond, and orange star respectively.}
\label{fig:c4_fig2}
\end{figure}

Figure~\ref{fig:c4_fig2} shows the 36 galaxies in the $z\sim2$ kinematic sample overlaid on the parent population of 557 $z\sim2$ LBGs dispersed on a $(U_n-\cal{R})$/$\cal{R}$ CMD adapted from Paper\,I.  We note that most of the $z\sim2$ kinematic sample is distributed toward the redder half of the CMD; 31 of 36 galaxies lie above the primary cut,.  and only five galaxies lie below it.  To a large degree, this tendency reflects the non-random selection bias in the source IFU studies, and the observational difficulty associated with obtaining high-quality IFU-based data for faint and/or compact sources.  For example, the early observations of FS09 targeted galaxies with `higher mass, spatially resolved velocity gradients, large velocity dispersions, or spatially extended emission', and WI15 reports specifically on a `high S/N disc sample' drawn from the larger \KMOS\ survey.  Due to the known relationship between \lya\ and $z\sim2$ LBG morphology \citep{Law2012c}, and that between rest-frame UV colour, \lya\ absorption/emission strength, and rotational dynamic support that we report herein, both of these selection biases inevitably result in samples that are redder than the average of the SFG population accessible by ground-based IFU spectrographs and our LBG parent sample.  That being said, the aims of this paper do not require that the kinematic galaxies sample the parent LBG population in an unbiased way.  It is sufficient that the kinematic sample spans a wide enough range in apparent magnitude and colour to identify a relationship with the broadband colour and magnitudes of LBGs, and that this relationship can be used to statistically segregate the galaxies by their \lya\ properties and general kinematic type.

This `red bias' of the $z\sim2$ kinematic sample notwithstanding, all but one of the 22 rotation-dominated galaxies lie above the primary cut, and all are above the dotted-dashed line, in the white and red regions that enclose $\gtrsim95$~percent of the aLBG population.  The `high-confidence' aLBG (red) region is free from contamination by non-merger dispersion-dominated galaxies, and only one such source (Q1217-BX95, see Section~\ref{sec:c4_elbgs}) is found above the primary cut.  Although few in number, all dispersion-dominated galaxies lie below the dashed line (in the white and blue regions) where $\gtrsim97$~percent of the eLBGs are located.  All but one galaxy (Q1623-BX502, see Section~\ref{sec:c4_elbgs}) below the primary cut are dispersion-dominated systems (including mergers).  Thus, as illustrated by the mean positions of the `rd' and `dd' sub-samples in Figure~\ref{fig:c4_fig2} (red circle and blue diamond respectively), the rotation- and dispersion-dominated subsets of the kinematic sample are coincident with, and may follow, the colour distributions of aLBGs and eLBGs on the CMD.  

\begin{figure}
\centering
\includegraphics[width=\columnwidth]{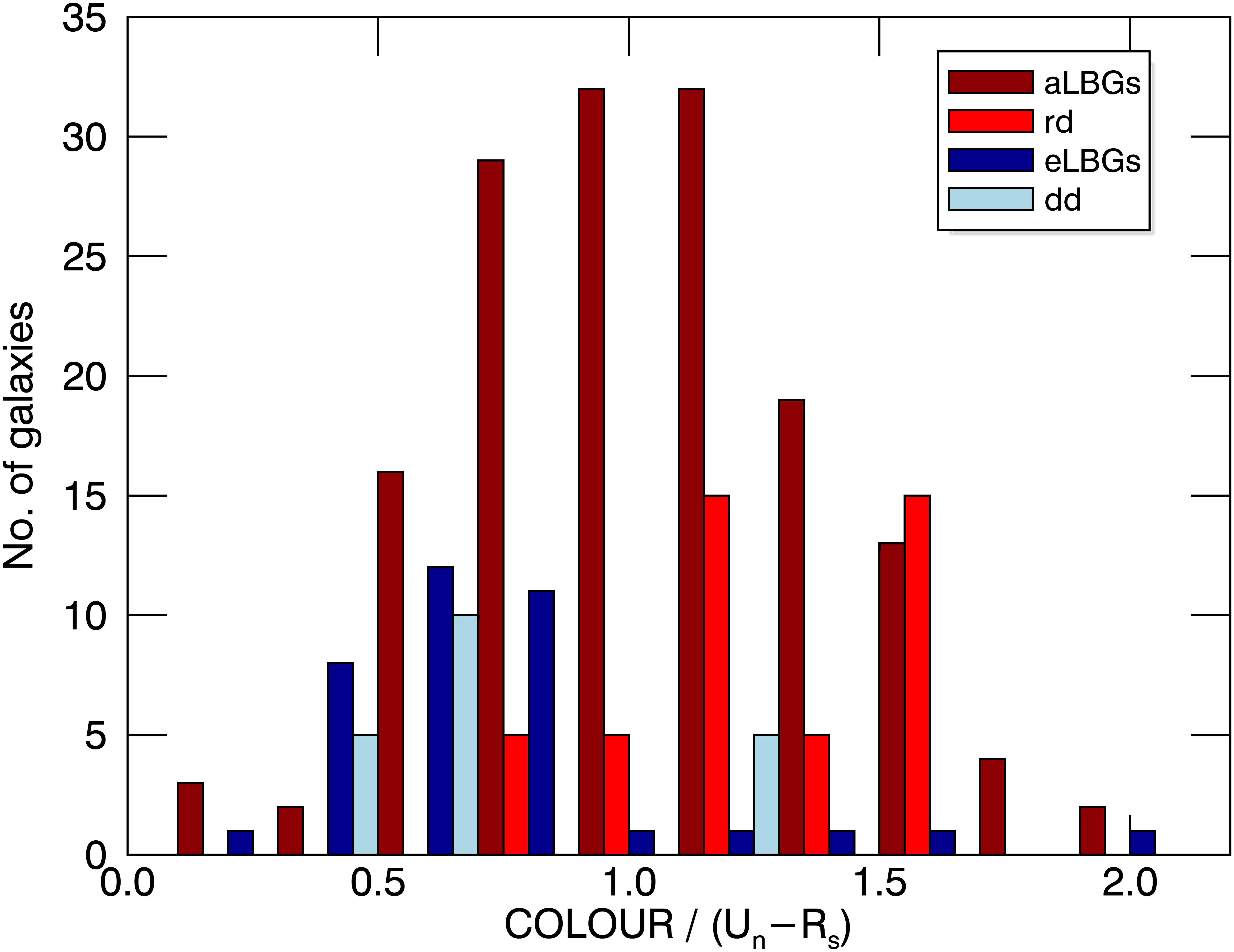}
\includegraphics[width=\columnwidth]{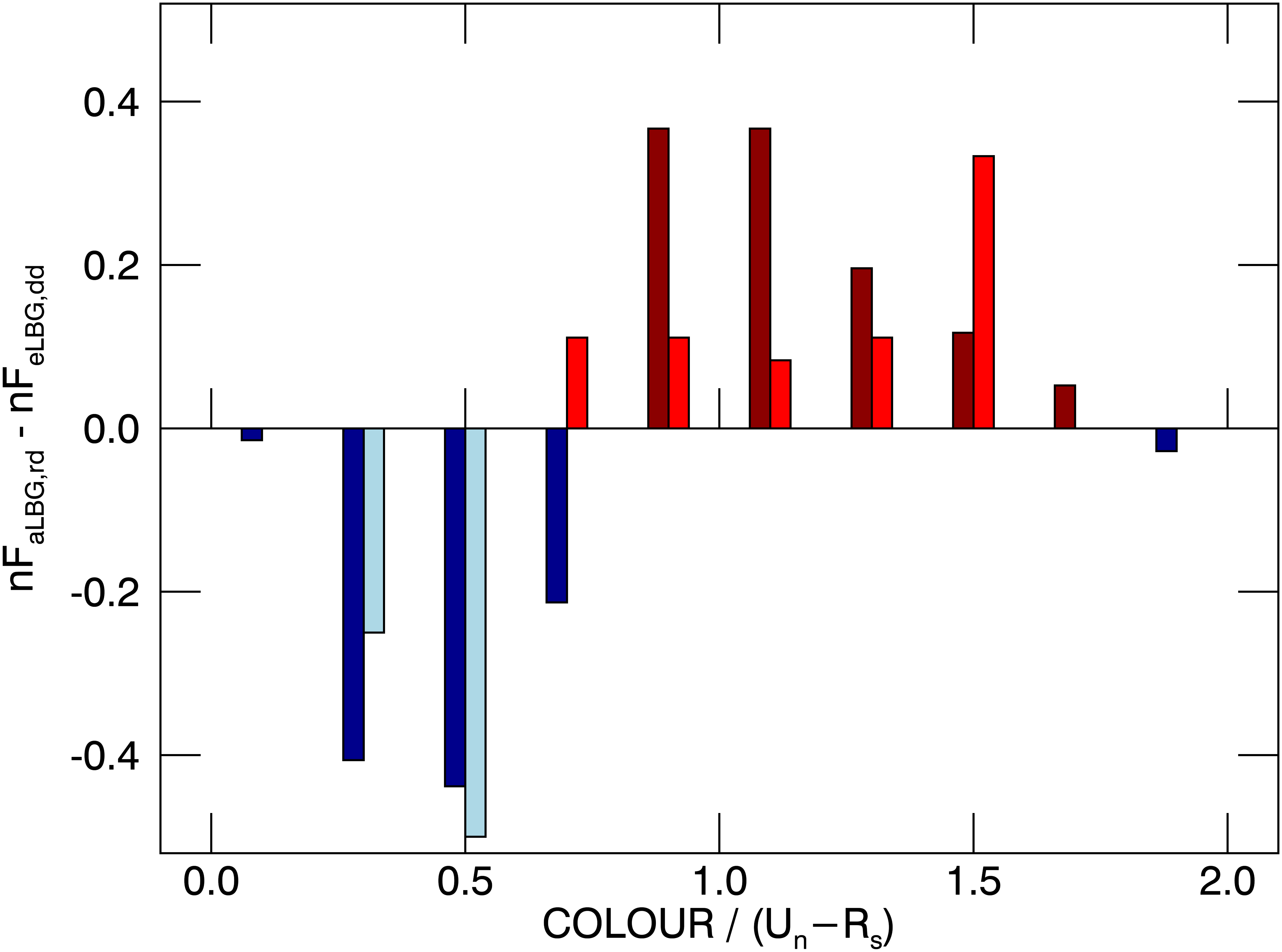}
\caption{Top: Histograms in $(U_n-\cal{R})$ colour of the aLBG, eLBG, `rotation-dominated' (rd) and `dispersion-dominated' (dd) sub-samples used in the KS tests.  Bottom: Histograms of the difference in normalised fraction of the same aLBG/eLBG and rd/dd sub-samples showing clear bifurcation in $(U_n-\cal{R})$ colour.}
\label{fig:c4_fig3}
\end{figure}

\begin{figure} 
\centering
\includegraphics[width=\columnwidth]{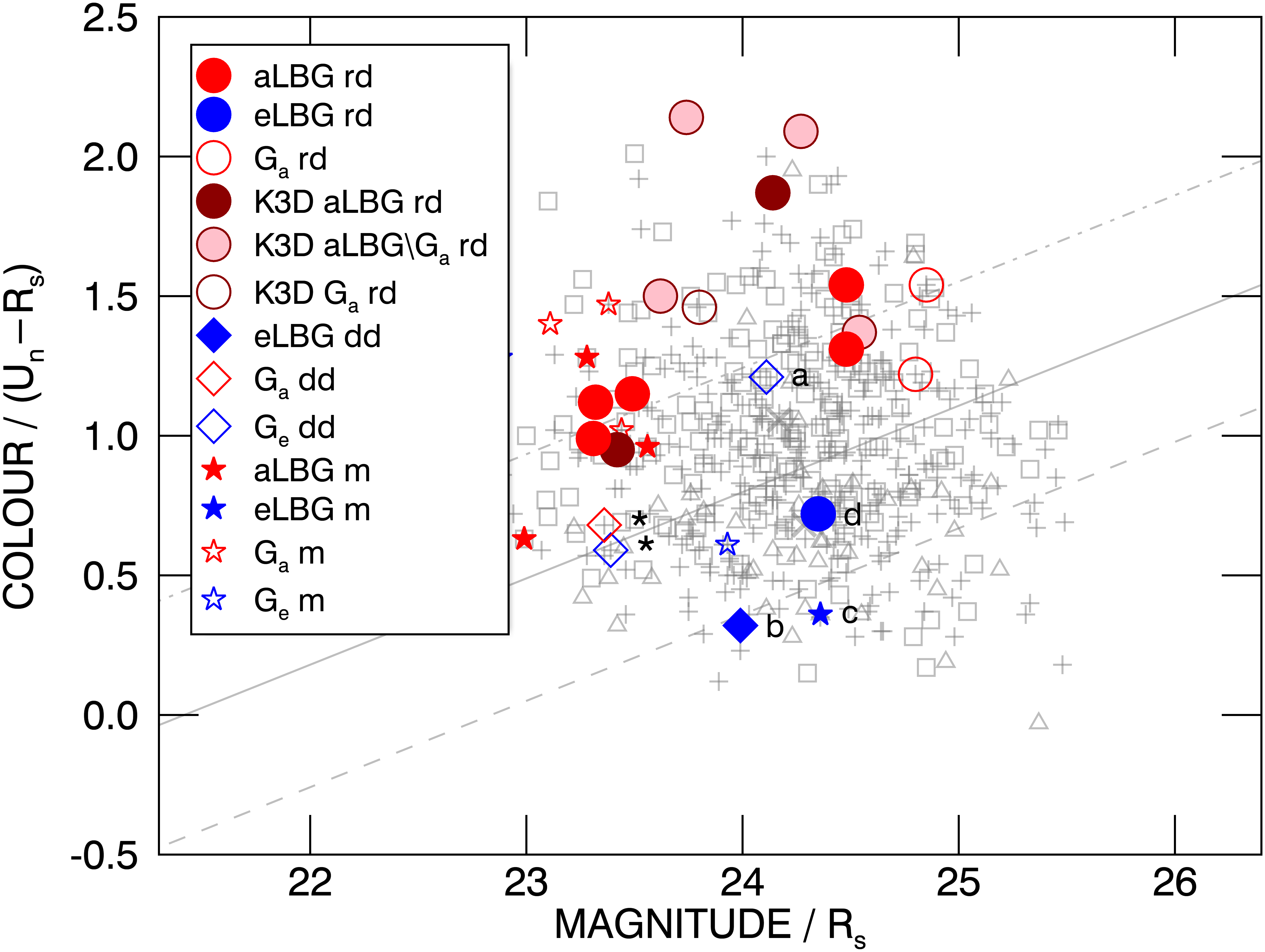}
\includegraphics[width=\columnwidth]{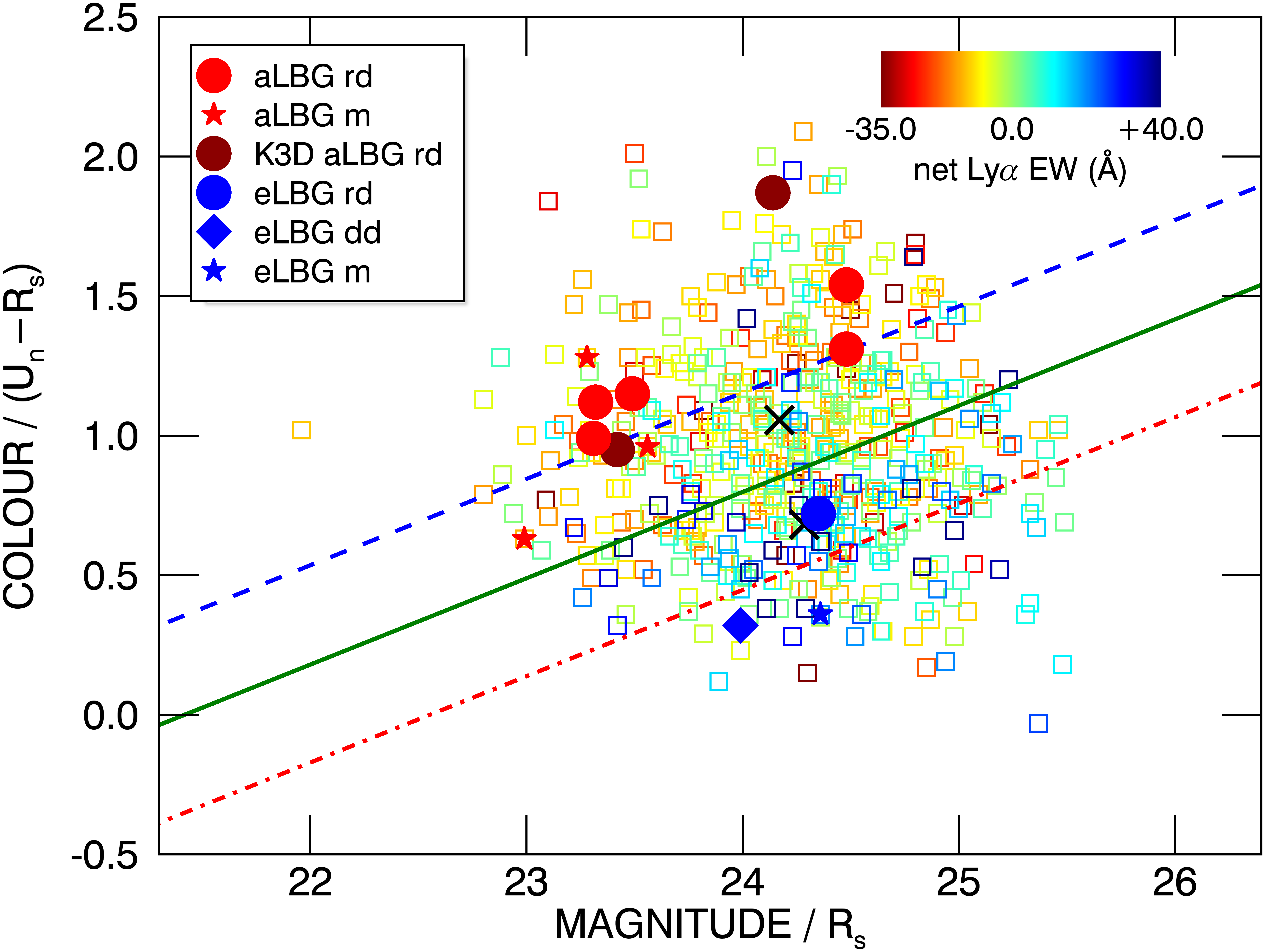}
\caption{Top: Similar to the right panel of Fig.~\ref{fig:c4_fig1} but for the $z\sim2$ kinematic sample plotted on a $(U_n-\cal{R})$ vs $\cal{R}$ CMD.  Members of the interacting close-pair identified by FS09 (Q2346-BX404 \& Q2346-BX405) are indicated by an asterisk ($^*$).  Other galaxies highlighted in Section~\ref{sec:c4_elbgs} are labelled as follows: (a) Q1217-BX95, (b) Q2343-BX418, (c) Q2343-BX660, and (d) Q1623-BX502. Bottom: Similar to the top panel, but with members of the kinematic sample that meet the criteria for classification as aLBGs and eLBGs overlaid on the parent $z\sim2$ LBGs (squares) colour-coded on a red-blue gradient according to their net \lya\ EW.  Black crosses indicate the mean colour and magnitude positions for the $z\sim2$ aLBG and eLBG distributions bisected by the solid green line (primary cut).  Dashed blue and dotted-dashed red lines indicate a 1-$\sigma$ dispersion in colour from the primary cut for the aLBG and eLBG distributions respectively, and define one choice of photometric selection criteria for the isolation of pure \lya-absorbing and \lya-emitting sub-samples as determined in Paper\,I.}  
\label{fig:c4_fig4}
\end{figure}

To test these propositions, we use the non-parametric two-sided Kolmogorov--Smirnov (KS) test to evaluate whether the distributions of the rotation and dispersion-dominated sub-samples in $(U_n-\cal{R})$ colour are statistically consistent with the null hypotheses that they are drawn from the $(U_n-\cal{R})$ colour distributions of the parent aLBG and eLBG populations.  For consistency in comparison, we limit the magnitude range of the underlying aLBGs and eLBGs to be the same as that spanned by the kinematic sample ($23.31 < \rm{M_{AB}} < 24.85$), and the kinematic sample to galaxies that derive from the UV-colour-selected $z\sim2$ LBG sample used to establish the photometric segregation and selection criteria described in Paper\,I, and shown on the CMDs in Figures~\ref{fig:c4_fig2}~\&~\ref{fig:c4_fig4}.  Histogram plots in $(U_n-\cal{R})$ colour of the sub-samples used in the KS test are shown in the top panel of Figure~\ref{fig:c4_fig3}.  The bifurcation in $(U_n-\cal{R})$ colour is more readily visualised in the lower panel of Figure~\ref{fig:c4_fig3} which shows colour histograms of the difference in normalised fraction of the same aLBG \& eLBG, and rd \& dd sub-samples (i.e., $\mathrm{nFrac}_{aLBG} - \mathrm{nFrac}_{eLBG}$ and $\mathrm{nFrac}_{rd} - \mathrm{nFrac}_{dd}$).

The KS tests yield a probability (p) of 0.969 that the dispersion-dominated galaxies in our kinematic sample and the underlying eLBG population derive from the same distribution of $(U_n-\cal{R})$ colours.  Moreover, we can reject with greater than 95\% confidence (p = 0.046) the null hypothesis that the dispersion-dominated sub-sample and aLBGs are drawn from the same colour distribution.  The analogous hypothesis that the rotation-dominated sub-sample and eLBGs share a common distribution of $(U_n-\cal{R})$ colours can be similarly rejected with high confidence (p = 0.0002), and there is a probability of p = 0.181 that they derive from the same distribution of aLBG colours.

The lower confidence of the  KS result for the rotation-dominated  sub-sample with respect to aLBGs can be plausibly explained in terms of the non-random selection bias in the kinematic sample described above.  Nevertheless, a KS test between the rotation and dispersion-dominated sub-samples indicates that we can reject with $\gtrsim$ 95\% confidence (p = 0.0425) -- and $\sim$ 98\% (p = 0.0203) if we include the \KMOS\ discs -- the null hypothesis that they are drawn from the same distribution in $(U_n-\cal{R})$ colour.

Sources classified as mergers will often contain two galaxies and will have brighter magnitudes compared to the non-merging fraction of the kinematic sample.  The mean $\cal{R}$-band magnitude of the ten merging systems is m($\cal{R}$)$_{merger}$ = 23.51 (orange star in Figure~\ref{fig:c4_fig2}), which is $\sim0.6$ mag brighter than the mean magnitude of the non-merging fraction (m($\cal{R}$)$_{non-merger}$ = 24.10).  Moreover, galaxies in the kinematic sample segregate in colour depending on their classification.  Mergers would statistically contain two galaxies roughly randomly selected from the aLBG, eLBG, G$_a$, and G$_e$ populations (modulo any environmental effects).  The mean $(U_n-\cal{R})$ colour (and standard deviation) of the mergers is 1.03 (0.38) compared to 1.43 (0.37) and 0.70 (0.37) for the rotation and dispersion-dominated sub-samples respectively.

The statistical evidence connects the kinematic properties of $z\sim2$ LBGs to \lya\ spectral type via their mutual association with rest-frame UV colour on the CMD.  We next look to investigate this trend directly using the 28 galaxies in the kinematic sample for which spectroscopically determined net \lya\ EWs are available.  The top panel of Figure~\ref{fig:c4_fig4} shows this sub-sample with their assigned \lya\ spectral types overlaid on the parent $z\sim2$ LBGs.  Seven out of eight rotation-dominated galaxies are aLBG or G$_a$ spectral types, and all seven \KMOS\ rotators are net \lya-absorbers (see Section~\ref{sec:c4_kmos}).  Three out of four dispersion-dominated galaxies are net \lya-emitters (eLBG or G$_e$ spectral types.  The lower panel of Fig.~\ref{fig:c4_fig4} shows just those members of the kinematic sample that meet the criteria for classification as aLBGs or eLBGs overlaid on the parent sample dispersed in colour-magnitude space with symbols colour-coded on a red-blue scale according to their measured net \lya\ EW as described in Paper\,I.  Seven of the 10 aLBGs are rotation-dominated and three are mergers.  All rotation-dominated aLBGs lie above the primary cut, with six of the seven above the dashed blue line in the `high-confidence' aLBG region.  The three eLBGs are located near or below the $z\sim2$ eLBG distribution mean, with two galaxies (including the dispersion-dominated Q2343-BX418) below the dotted-dashed red line in the `high-confidence' eLBG region.

Overall, not only do the rotation and dispersion-dominated sub-samples lie on the CMD in a way that is statistically associated with the distribution in colour of the \lya-absorbing and \lya-emitting populations, they have spectroscopically determined \lya\ spectral types that are consistent with this trend.  Moreover, reviewing both panels of Figure~\ref{fig:c4_fig4}, the broadband photometric selection criteria select nearly pure samples of rotation- and dispersion-dominated systems and potentially 100\% pure selection for galaxies meeting our net \lya\ EW criteria.  

\subsubsection{eLBGs and dispersion-dominated galaxies}
\label{sec:c4_elbgs}

The absence of galaxies with large, disc-like rotating structures in the \lya-emitting half of the CMD, and the fact that the `high-confidence' aLBG region is almost completely devoid of net \lya-emitting and/or dispersion-dominated sources, suggest that strong \lya\ emission is linked to dispersion-dominated kinematics.  The strength of such a claim is, however, challenged by the relative scarcity of faint, blue \lya-emitting galaxies in our sample.  In this regard, a closer examination of the eLBG and dispersion-dominated galaxies in the kinematic sample is instructive. 

The four dispersion-dominated galaxies in the $z\sim2$ kinematic sample (diamond symbols in Figure~\ref{fig:c4_fig2} and the top panel of Figure~\ref{fig:c4_fig4}) have a mean net \lya\ EW of $+$14.4 \AA.  Of these, three are net \lya-emitters (eLBG or G$_e$ spectral types).  The exception is Q2346-BX405 which has a net \lya\ EW of $-8.61$\,\AA\ (G$_a$ spectral type).  While FS09 ascribe a `dispersion-dominated' classification to Q2346-BX405, they note that it also has some kinematic features that are consistent with disc rotation.  Moreover, they point out that Q2346-BX404 and Q2346-BX405 are an interacting pair -- a feature that could result in the partial disruption of rotation kinematics and/or a subsequent classification as a dispersion-dominated system.  In any event, Q2346-BX405 and its partner Q2346-BX404 (marked by asterisks in the top panel of Figure~\ref{fig:c4_fig4}) lie near the centre of the CMD at positions we find to be typical of mergers, and would be excluded from selection as aLBGs or eLBGs in our photometric method.

The other dispersion-dominated galaxy of interest within our framework is Q1217-BX95 (galaxy `a' in the top panel of Figure~\ref{fig:c4_fig4}).  While its location on the CMD is within the eLBG colour--magnitude distribution, it has a colour ($(U_n-\cal{R})$ = 1.2) that is more typical of a rotation-dominated net \lya-absorber.  Q1217-BX95 is relatively massive ($\mathrm{log}({M}_{\star }/{M}_{\odot })=10.08$), but is compact ($R_e$ = 0.6 kpc), has low gas-phase metallicity ($12 + \mathrm{log}({O}/{H}) = 8.15 \pm 0.02$), and displays a $5\sigma$ detection of the auroral [O{\footnotesize III}]$\lambda$4363 emission line\footnote{Gas-phase metallicities supplied by C.\ Steidel (priv.\ comm.) determined using the O3N2 index and the calibration described by \citet{Steidel2016} and \citet{Strom2017}.  Values are quoted using the scale of \citet{Asplund2009} in which solar metallicity has the value $12 + \mathrm{log}({O}/{H}) = 8.69$}.  On the basis of this ensemble of properties, and a net \lya\ EW of $+$10.2\,\AA, we would predict Q1217-BX95 to have dispersion-dominated kinematics, consistent with the classification for this galaxy reported by LA09.  We note that photometric errors deriving from poor seeing in the Q1217 field images ($1.56\dblprime$ in $U_n$) may be responsible for the $(U_n-\cal{R})$ colour of this galaxy (C.\ Steidel, priv.\ comm.).

The three spectroscopic eLBGs in our kinematic sample (Q2343-BX418, Q2343-BX660 and Q1623-BX502) are the three sources furthest below the primary cut on the CMD, and lie close to the eLBG mean (see bottom panel of Figure~\ref{fig:c4_fig4}).  They are, however, kinematically diverse.  The dispersion-dominated Q2343-BX418 (galaxy `b' in the top panel of Figure~\ref{fig:c4_fig4}) has low stellar mass ($\mathrm{log}({M}_{\star }/{M}_{\odot })= 9.0$), low metallicity ($12 + \mathrm{log}({O}/{H}) = 7.9 \pm0.2$), is very blue ($(U_n-\cal{R})$ = 0.32), very compact ($R_e$ = 0.8 kpc), and has strong \lya\ emission (net \lya\ EW = $+$53.5 \AA).  Although having kinematic structure consistent with merging compact galaxies (LA09), Q2343-BX660 (galaxy `c' in the top panel of Figure~\ref{fig:c4_fig4}) is otherwise similar to Q2343-BX418; it has relatively low stellar mass ($\mathrm{log}({M}_{\star }/{M}_{\odot })= 9.9$), low metallicity ($12 + \mathrm{log}({O}/{H}) = 7.99 \pm0.05$), compact morphology ($R_e$ = 1.6 kpc), very blue colour ($(U_n-\cal{R})$ = 0.36), and is a strong \lya\ emitter (net \lya\ EW = $+$20.4\,\AA).

With a net \lya\ EW of $+$28.5\,\AA, and a `rotation-dominated' kinematic classification derived from the deep AO-assisted observations of FS18, Q1623-BX502 (galaxy `d' in the top panel of Figure~\ref{fig:c4_fig4}) appears to be inconsistent with our proposition that large net \lya\ EW is a useful predictor of dispersion-dominated kinematics.  Although its magnitude and redder colour $(U_n-\cal{R})$ = 0.72) than the other eLBGs place it in the `white strip' on the CMD that contains both aLBGs and eLBGs, its relatively low stellar mass ($\mathrm{log}({M}_{\star }/{M}_{\odot })= 9.4$), compact morphology ($\rm{R_e} = 1.1 \pm 0.7$kpc), low gas phase metallicity ($12 + \mathrm{log}({O}/{H}) = 8.09$), are typical of what we would predict for a dispersion-dominated galaxy with such strong \lya\ emission.  Earlier reports based on seeing-limited and AO observations classify Q1623-BX502 as `dispersion-dominated' \citep[FS09, LA09 and ][]{Newman2013} and the ratios ${\rm{\Delta}}{v}_{\mathrm{obs}}/2{\sigma }_{\mathrm{tot}}$ ($0.50 \pm 0.16$) and ${V}_{\mathrm{rot}}/{\sigma}_{0}$ (${2.0}_{-0.8}^{+1.5}$) for Q1623-BX502 reported by FS18 are equivalent (within the stated uncertainties) to the threshold values (0.4 and $\sqrt{3.36} \approx 1.83$ respectively) used by FS09 and FS18 to classify galaxies as either rotation or dispersion-dominated (see Section~\ref{sec:c4_kinemetry}).  Thus, Q1623-BX502 is a borderline case in terms of its kinematics, and, by its position on the CMD, would not be selected by our kinematic broadband photometric criteria (see Paper\,I).

To summarise, although some of the \lya-emitting sources in the $z\sim2$ kinematic sample are not classified as `dispersion-dominated' in the source IFU studies, they are characteristically compact, blue in colour, have relatively low stellar mass, low gas-phase metallicity, and highly disordered kinematics in which the rotation signature (if any) is weak and more typical of low angular momentum `orbital-type', rather than the strong `disc-like' rotation characteristic of the \lya\ absorbers.  We note that those cases that are borderline or complex do not meet our broadband photometric criteria, and would not be selected.

\subsection{Kinematics vs net \lya\ EW -- Direct Comparisons}
\label{sec:c4_direct}

\subsubsection{Kinematic classifications vs net \lya\ EW}
\label{sec:c4_class}

With even a naive application of the kinematic classifications derived from the literature, there is a clear bifurcation in the average \lya\ spectral properties of the non-merger rotation and dispersion-dominated populations in the $z\sim2$ and $z\sim3$ kinematic samples.  The mean net \lya\ EW for rotation-dominated galaxies in the $z\sim2$ sample is $-$7.0 \AA, and $+$14.4 \AA\ for the dispersion-dominated sub-sample.  (cf.\ $+$1.1 and $+$19.6~\AA\ for the $z\sim3$ rotating/rotation-dominated and not-rotating/dispersion-dominated sub-samples respectively).  A two-sided KS test on a combined ($z\sim2$ + $z\sim3$) sample comprising eighteen rotation- and 14 dispersion-dominated galaxies rejects with $\sim$99\% confidence (p = 0.012) the null hypothesis that the rotation-and dispersion-dominated galaxies were drawn from the same distribution of net \lya\ EW values.

\subsubsection{Quantitative kinematics versus \lya}
\label{sec:c4_kinemetry}

We now look to investigate the relationship between net \lya\ EW and galaxy kinematics using quantitative parameters derived from the IFU kinematic maps of the respective source studies in addition to the classifications that we have used up to this point.  In doing so, we note that methods to measure rotation velocity and global velocity dispersion differ between surveys, Although we have taken care to compare like with like as much as possible, homogenising these methods is beyond the scope of this paper, and we acknowledge that this is a potential source of systematic errors.

For this purpose, we invoke the ratio \vobs\ used by FS09 (and with modified notation by FS18) where ${v}_{\mathrm{obs}}/2$ is half the maximum observed \halpha\ velocity gradient across the source, and ${\sigma }_{\mathrm{int}}$ (${\sigma }_{\mathrm{tot}}$ in FS18) is the integrated velocity dispersion derived from the linewidth of the \halpha\ emission in the spatially collapsed object spectrum.  We note that ${v}_{\mathrm{obs}}/2$ and ${\sigma}_{\mathrm{int}}$ are equivalent to the shear velocity (${v}_{\mathrm{shear}}$) and net velocity dispersion (${\sigma}_{\mathrm{net}}$) respectively, in the nomenclature of LA09.  We similarly equate for the purpose of this analysis, the observed rotation velocity (V$_{\mathrm{obs}}$) and velocity dispersion ($\sigma_{\mathrm{obs}}$) of TU17 with ${v}_{\mathrm{obs}}/2$ and ${\sigma}_{\mathrm{int}}$ respectively.

As a quantity derived from the observed velocity field, ${v}_{\mathrm{obs}}/2$ is related to the actual rotation velocity by a factor of $1/\sin(i)$, where $i$ is the galaxy inclination angle, and ${\sigma }_{\mathrm{int}}$ incorporates contributions to the integrated linewidth from both large-scale velocity gradients and intrinsic (random) gas motions.  Nevertheless, \vobs\ has proven to be a useful (if approximate) probe of the nature of the dynamical support (FS09), and provides here a ready means to compare kinematic results from different studies.  Values of ${v}_{\mathrm{obs}}/2$, ${\sigma}_{\mathrm{int}}$ and the ratio \vobs\ for a subset of 21 $z\sim2$ and five $z\sim3$ galaxies in our kinematic samples derived from their respective source studies are listed in Tables~\ref{tab:c4_table3}~\&~\ref{tab:c4_table5}. 

\begin{figure}
\centering
\includegraphics[width=\columnwidth]{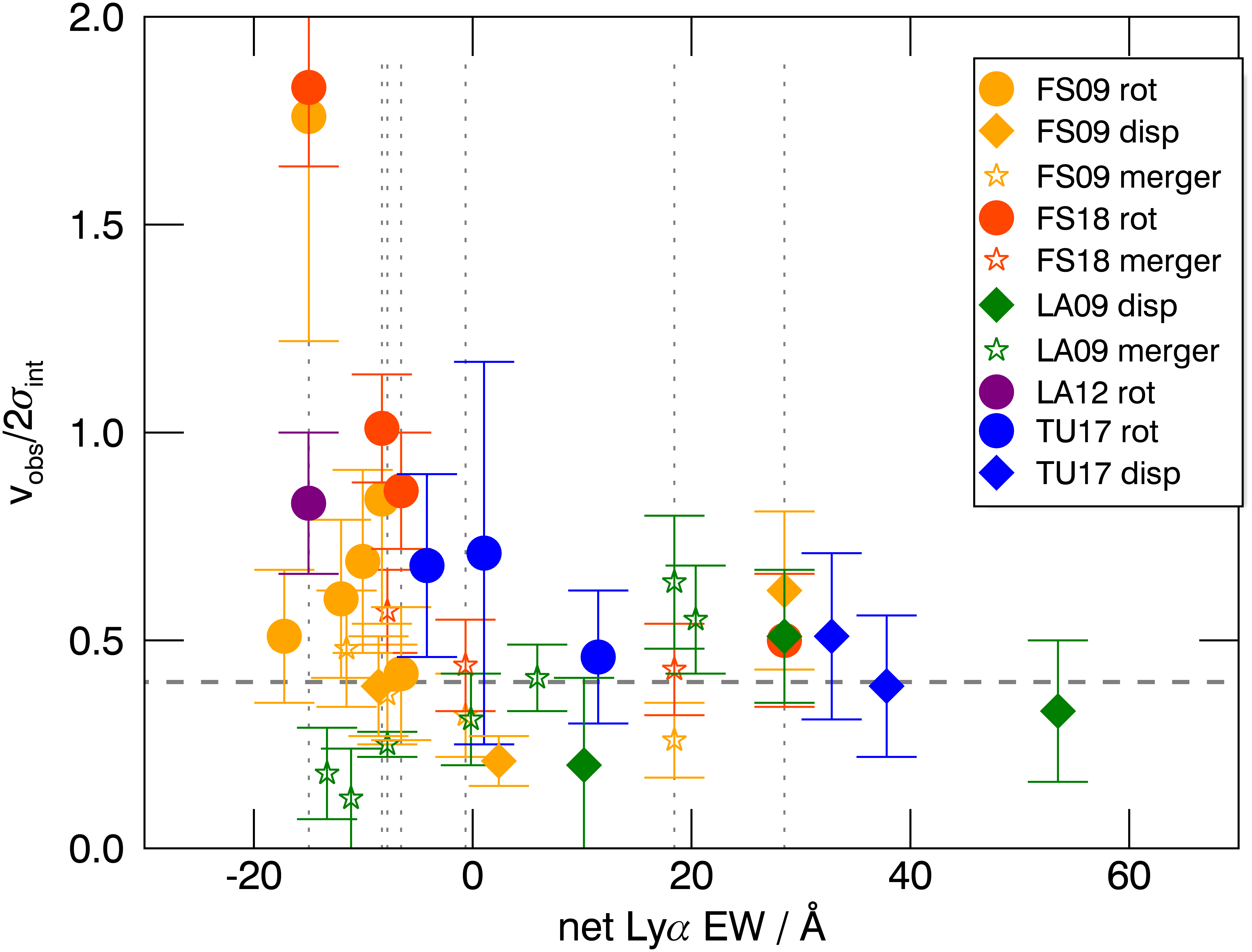}
\caption{\lya\ versus the ratio of observed rotation velocity to integrated velocity dispersion (\vobs) for UV-colour-selected $z\sim2-3$ LBGs from FS09 (orange), LA09 (green), LA12 (purple), FS18 (dark orange) and TU17 (blue) plotted as a function of net \lya\ EW and kinematic classification.  Uncertainties are as given in the source publications or, for Q2343-BX442 (LA12) as supplied by D.\ Law (priv.\ comm.).  Galaxies with \vobs\ estimates from multiple sources are aligned in net \lya\ EW and indicated with vertical dotted lines.  Rotation-dominated galaxies are shown as circles, dispersion-dominated systems as diamonds, and mergers as stars.  The horizontal dashed line indicates the threshold value (\vobs\ = 0.4) used by FS09 to classify galaxies as either rotation- or dispersion-dominated.}  
\label{fig:c4_fig5}
\end{figure}

Figure~\ref{fig:c4_fig5} shows \vobs\ (with supplied uncertainties) for these galaxies plotted versus net \lya\ EW.  Galaxies that are common between the different source studies and observation modes, and for which we have multiple estimates of \vobs, lie on vertical dotted lines to identify them for clarity.  LA09 conclude that observational bias is an unsatisfactory explanation for differences in the global kinematic properties of the Keck/OSIRIS (LA09) and VLT/SINFONI (FS09) samples, and similar kinematics were found for the galaxies that were successfully observed by both surveys.  There is also excellent agreement between the kinematic and morphological properties of Q1623-BX502 determined by both seeing-limited (FS09) and AO-assisted observations (FS09, LA09, FS18).  Moreover, all the galaxies plotted in Figure~\ref{fig:c4_fig5} are drawn from the related $z\sim2$ and $z\sim3$ UV-colour-selected catalogs of \citet{Steidel2003, Steidel2004} and are, therefore, comparable in terms of their intrinsic properties (see Section~\ref{sec:c4_overview}).  We can be confident, therefore, that the dispersion versus net \lya\ EW shown in Figure~\ref{fig:c4_fig5} derives from a genuine range in the physical properties of the galaxies (see Section~\ref{sec:c4_others}), and that the kinematics probed by our sample, from the multiple surveys, reliably reflect the true continuum of kinematic character ranging from genuinely dispersion-dominated to rotationally supported systems.  

The horizontal dashed line in Figure~\ref{fig:c4_fig5} at \vobs\ = 0.4 marks the threshold for the classification of galaxies as rotation or dispersion-dominated as established by FS09 for both seeing-limited and AO-assisted observations at typical spatial resolution.  Where this criterion is applied to mergers, ${v}_{\mathrm{obs}}/2$ can be interpreted as the projected `orbital velocity' of the system.

The upper right-hand corner of Figure~\ref{fig:c4_fig5} is unpopulated; there are no net \lya-emitters with strong rotational signatures.  All galaxies with strong rotational support are \lya\ absorbers and occupy the upper left corner of the plot.  The lower left-hand corner (\vobs $\lesssim 0.4$ and net \lya\ EW $\lesssim$ 0 \AA) is dominated by merging systems.  The only salient exceptions to these trends are: Q2346-BX405 (net \lya\ EW $= -8.6$), which is classified as dispersion-dominated (\vobs$= 0.21 \pm 0.06$), but shows kinematic features consistent with disc rotation and is part of an interacting pair (FS09); and Q1623-BX502 (net \lya\ EW = $+$28.5~\AA) and SSA22a-D3 (net \lya\ EW = $+$11.47~\AA), both of which have low angular momentum ($v_{\mathrm{obs}}/2 = \sim$40 and 53\kms\ respectively) and are borderline for classification as rotation- or dispersion-dominated.

Parameterisation of the kinematics of our sample via \vobs\ allows us to statistically test the strength of the association between galaxy kinematics and net \lya\ EW independent of the inherited kinematic classifications.  For galaxies with multiple estimates of \vobs, we calculate average values and derive a sample of 17 discrete non-merger galaxies.  A Spearman rank order correlation test on this sub-sample yields a coefficient ($\rho$) of $\sim$0.577 and a probability (p) of 0.0154 ($\sim2.2\sigma$ significance) that this moderate to strong degree of association could arise from an uncorrelated sample.  

While \vobs\ is a useful empirical measure of rotational dynamic support, a more robust and physically intuitive parameter for this purpose is the ratio \vrsO\ invoked by FS18, where $v_{\mathrm{rot}}$ is the intrinsic rotation velocity corrected for beam smearing and inclination angle according to:

\begin{equation} 
\mbox{$v_{\mathrm{rot}}$} \cdot \mbox{$\sin(i)$} ~ = ~ \mbox{$C_{\mathrm{PSF,}v}$} \cdot \mbox{$v_{\mathrm{obs}}/2$} 
\label{eq:c4_v_rot}
\end{equation} 

\noindent and $\sigma_{\mathrm{0}}$ is the intrinsic local velocity dispersion determined from correction of the observed local velocity dispersion ($\sigma_{\mathrm{0obs}}$) for beam smearing according to: 

\begin{equation}
\mbox{$\sigma_{\mathrm{0}}$} ~ = ~ \mbox{$C_{\mathrm{PSF,}\sigma}$} \cdot \mbox{$\sigma_{\mathrm{0obs}}$,} 
\label{eq:c4_sigma_0} 
\end{equation} 

\noindent where $\sin(i)$ is the correction for inclination of a galaxy from the plane of the sky, and $C_{\mathrm{PSF,}v}$ and $C_{\mathrm{PFS,}\sigma}$ are the beam-smearing correction factors for rotation velocity and velocity dispersion respectively, in the rotating-disc framework of FS18.

For the SINS/zC-SINF galaxies of FS09 and FS18, and Q2343-BX442 from LA12, we use the $v_{\mathrm{rot}}$, $\sigma_{\mathrm{0}}$ and \vrsO\ values as published.  Applying Equation~\eqref{eq:c4_v_rot} to the LA09 galaxies, we use the average correction factor of the FS18 sample (i.e., $C_{\mathrm{PSF,v}} = 1.3$) in a manner similar to \citet{Tacconi2013}, and $\sin(i) = \pi/4$, the mean inclination angle correction for a distribution of randomly inclined discs (LA09) to estimate $v_{\mathrm{rot}}$.  For the purposes of this comparison, $\sigma_{\mathrm{mean}}$ in LA09 approximates to $\sigma_{\mathrm{0obs}}$ in FS18.  Following Equation~\eqref{eq:c4_sigma_0}, we apply to the LA09 sample a correction of $C_{\mathrm{PSF},\sigma} = 0.85$, the average correction factor for the FS18 sample.  Uncertainties are propagated from the LA09 estimates for $v_{\mathrm{shear}}$ and $\sigma_{\mathrm{mean}}$ and added in quadrature for the ratio \vrsO.  

\begin{figure*}
\centering
\includegraphics[width=0.45\columnwidth]{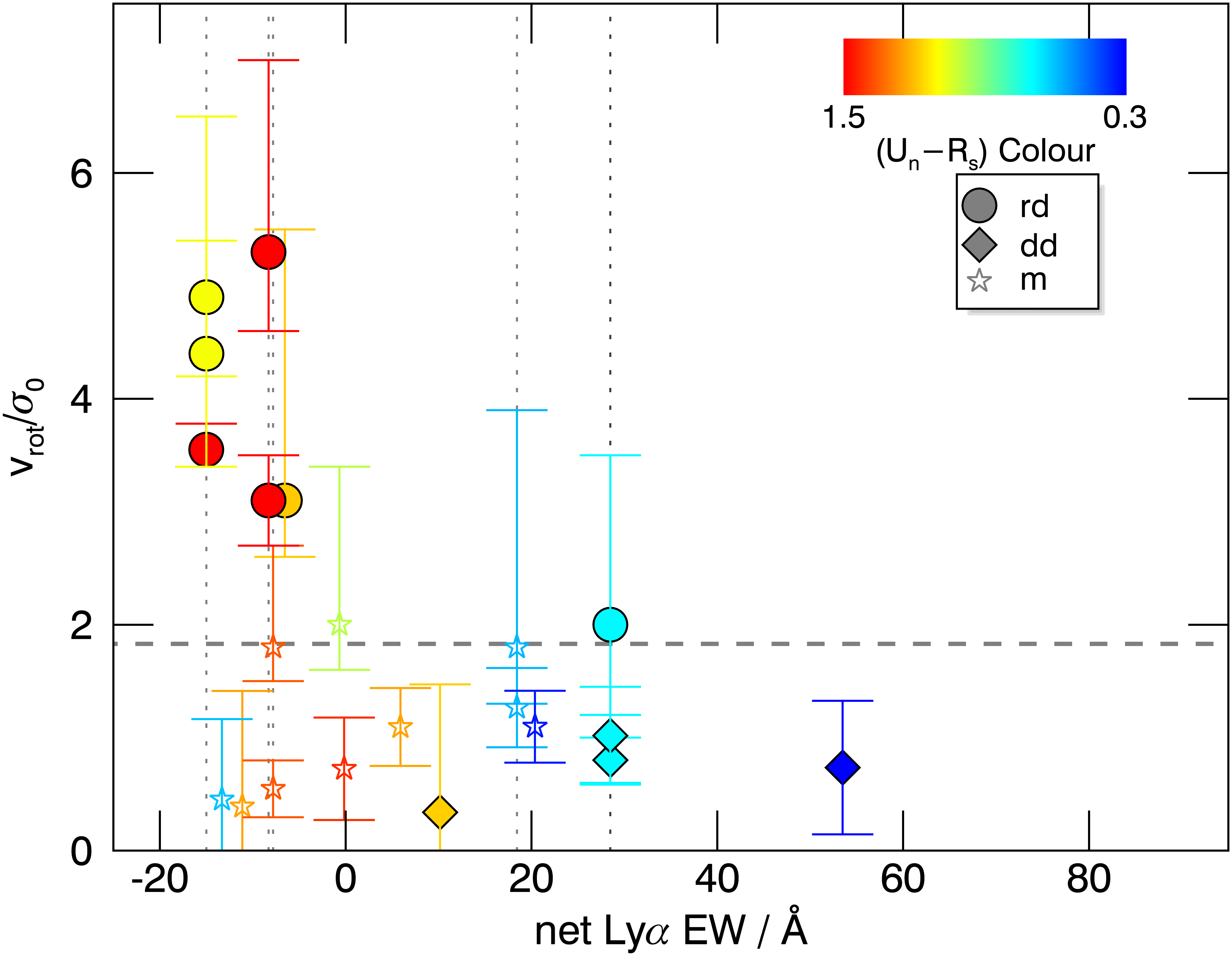}
\includegraphics[width=0.45\columnwidth]{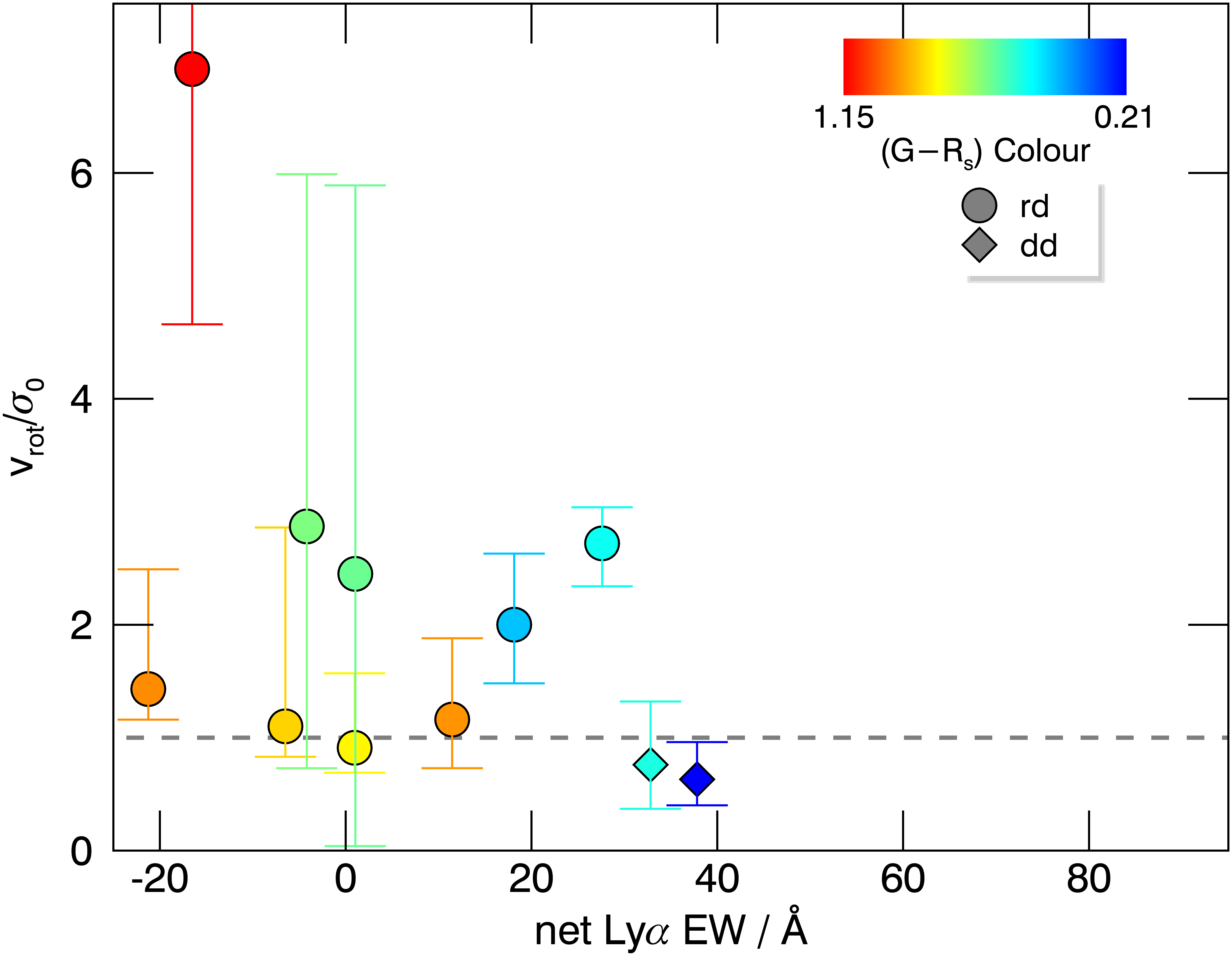}
\caption{Ratio of intrinsic rotation velocity to intrinsic velocity dispersion (\vrsO) for $z\sim2$ (left) and $z\sim3$ (right) LBGs plotted as a function of net \lya\ EW, rest-frame UV colour, and kinematic classification.  Rotation-dominated (rd) galaxies are shown as circles, dispersion-dominated (dd) systems as diamonds, and mergers (m) as stars.  Symbols are colour-coded on a blue/green/red gradient according to their rest-frame UV colours from $(U_n-\cal{R})$ = 0.3 to 1.5, and from $(G-\cal{R})$ = 0.21 to 1.15 for the $z\sim2$ and $z\sim3$ samples respectively.  The horizontal dashed lines mark the thresholds used by FS18 ($z\sim2$) and TU17 ($z\sim3$) to classify galaxies as either rotation- or dispersion-dominated.  Vertical dotted lines indicate galaxies with kinematic data from multiple surveys. Dynamical support due to rotation (as measured by \vrsO) correlates strongly with net \lya\ EW and rest-frame UV colour as predicted from the CMD results described in Sections~\ref{sec:c4_z3lbgs}~\&~\ref{sec:c4_z2lbgs}.}  
\label{fig:c4_fig6}
\end{figure*}

For the $z\sim3$ galaxies, we similarly extract from the source references values for intrinsic rotation velocity (V$_{\mathrm{max}}$ in GN11 and V$_{\mathrm{C}}$ in TU17) and intrinsic velocity dispersion ($\sigma_{\mathrm{int}}$ in both GN11 and TU17), and equate these to ${v}_{\mathrm{rot}}$  and ${\sigma}_{\mathrm{0}}$ respectively for the purposes of this analysis.  Values of $v_{\mathrm{rot}}$, $\sigma_{\mathrm{0}}$ and \vrsO\ used to construct Figure~\ref{fig:c4_fig6} are listed with supplied or derived uncertainties, and associated net \lya\ EWs, in Tables~\ref{tab:c4_table3}~\&~\ref{tab:c4_table5}.

\begin{figure}
\centering
\includegraphics[width=\columnwidth]{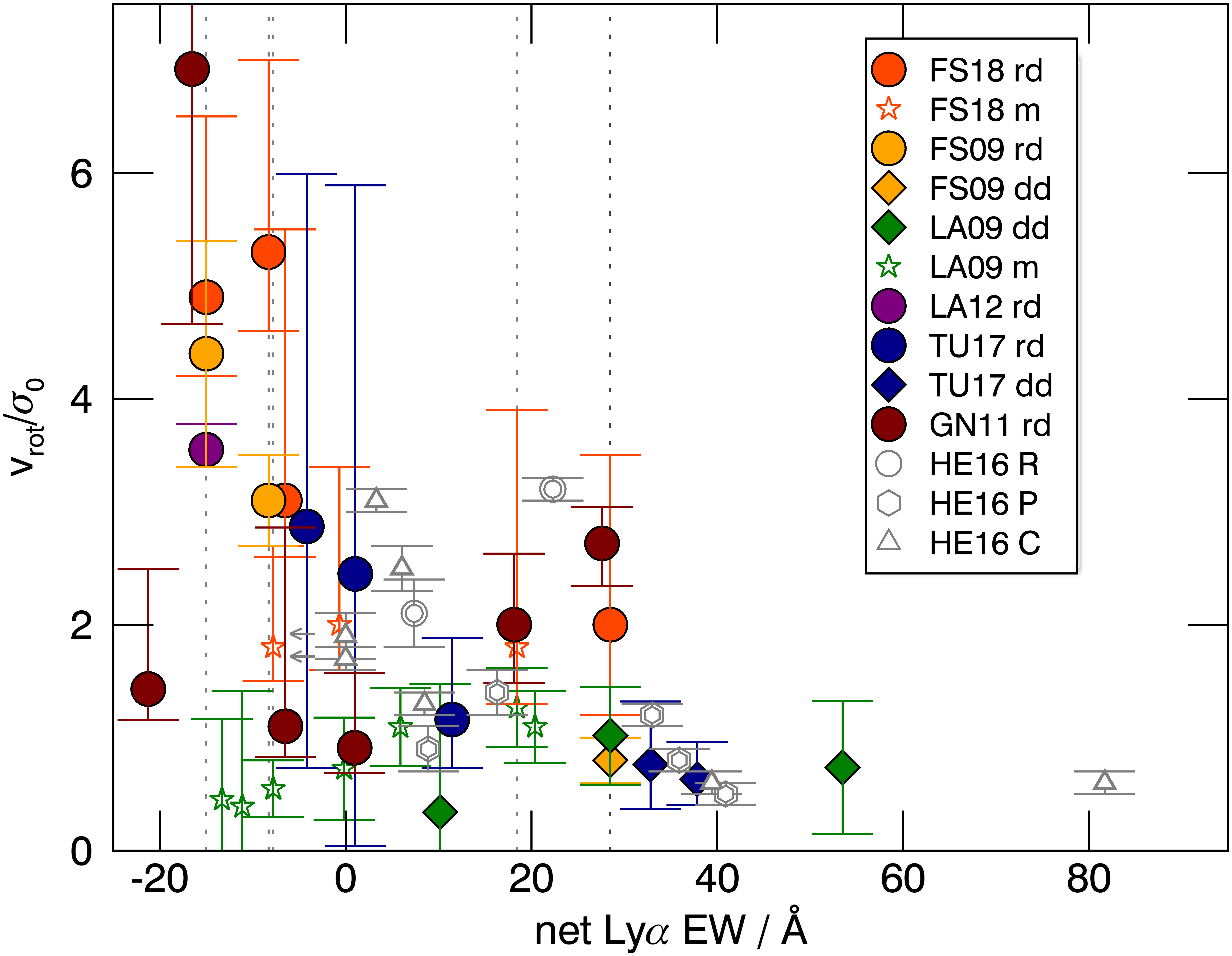}
\caption{Same as Figure~\ref{fig:c4_fig6} but with the $z\sim2$ and $z\sim3$ samples plotted together and colour-coded according to their respective source surveys.  Also shown are the 14 $z\sim0.03$ LARS galaxies (grey/white symbols) plotted using kinematic and \lya\ EW data from \citet{Herenz2016}.  LARS galaxies classified as `rotating discs' (R) are plotted as circles, `perturbed rotators' (P) as hexagons, and galaxies with `complex kinematics (C) as triangles.  The two LARS galaxies listed by \citet{Herenz2016} with a \lya\ EW of zero (LARS04 \& LARS06) are net absorbers at all apertures within the LARS field of view.  Grey arrows indicate the direction that these galaxies would move on the plot if this net absorbing character was reflected in the quoted \lya\ EWs.  (see Section~\ref{sec:c4_lowz}).} 
\label{fig:c4_fig6c}
\end{figure}

Figure~\ref{fig:c4_fig6} shows the ratio of intrinsic rotation velocity to intrinsic velocity dispersion (\vrsO) thus derived for subsets of our kinematic samples, plotted as a function of net \lya\ EW and rest-frame UV colour, i.e., $(U_n-\cal{R})$ for $z\sim2$ (left) and $(G-\cal{R})$ for $z\sim3$ (right).  Figure~\ref{fig:c4_fig6c} shows the \vrsO\ versus net \lya\ EW relationship for the $z\sim2$ and $z\sim3$ samples with galaxies colour-coded according to their source survey.  A Spearman rank correlation test for the \vrsO\ versus net \lya\ EW relationship for a combined non-merging subset of 18 $z\sim2$ and $z\sim3$ galaxies gives a $\rho$-value of -0.69, and a p-value of 0.0014 that allows us to reject the null hypothesis that there is no association between \vrsO\ and net \lya\ EW with $\gtrsim$ 99.8\% ($\sim3\sigma$) confidence.

The $z\sim2$ subset of strongly rotating `disc-like' sources (\vrsO$\gtrsim3$) have the most negative values of net \lya\ EW, and exclusively populate the top left corner of the left panel in Figure~\ref{fig:c4_fig6}, significantly above the horizontal dashed line at \vrsO$= \sqrt{3.36}$ ($\sim1.83$) that corresponds to the point above which rotation starts to dominate over velocity dispersion in the dynamical support of turbulent discs (FS18).  Similarly, non-merger galaxies with the lowest values of \vrsO have high net \lya\ EWs and lie toward the bottom right.  Mergers are confined to the bottom left corner of the plot.  In general, there is an increase in the dispersion of the sample in the \vrsO\ dimension relative to the \vobs\ case for the $z\sim2$ sample (cf.\ Figure~\ref{fig:c4_fig5})  indicating that as the rigour of the kinematic analysis is increased, the relationship between net \lya\ EW and the degree of rotational dynamic support is strengthened.

Differences in the details of the colour-colour selection criteria of the parent photometric catalogues (see Paper\,I) preclude direct comparison of the colour dispersion of the $z\sim2$ and $z\sim3$ kinematic samples.  However, in both redshift ranges (see Figure~\ref{fig:c4_fig6}), there is a clear trend from red to blue with decreasing rotational dynamic support (as measured by \vrsO) that provides quantitative verification, at a galaxy-by-galaxy level, of the statistical relationship between kinematic type, rest-frame UV colour, and net \lya\ EW, as described in Sections~\ref{sec:c4_z3lbgs}~\&~\ref{sec:c4_z2lbgs}.

\begin{table*}
\centering
\caption{Kinematic and \lya\ Properties of UV Colour-Selected $z\sim2$ LBGs}
\label{tab:c4_table3}
\scalebox{0.82}{%
\begin{threeparttable}
\begin{tabular}{ccccccccccc}  
\toprule
\thead{ID} & 
\thead{${v}_{\mathrm{obs}}/2$\tnote{a} \\ (km\,s$^{-1}$)} & 
\thead{${\sigma}_{\mathrm{int}}$\tnote{b} \\ (km\,s$^{-1}$)} & 
\thead{${v}_{\mathrm{obs}}/2{\sigma}_{\mathrm{int}}$\tnote{c}} & 
\thead{${v}_{\mathrm{rot}}$\tnote{d} \\ (km\,s$^{-1}$)} & 
\thead{${\sigma}_{\mathrm{0}}$\tnote{e} \\ (km\,s$^{-1}$)} & 
\thead{${v}_{\mathrm{rot}}/{\sigma}_{\mathrm{0}}$\tnote{f}} & 
\thead{net \lya\\ EW (\AA)} &  
\thead{Kin.\ \tnote{g}\\ class.} & 
\thead{Obs.\tnote{h} \\ mode} & 
\thead{Ref. \tnote{i}} \\
\midrule
HDF-BX1564 &  $12^{+12}_{-12}$ & $103^{+14}_{-14}$ & $0.12^{+0.12}_{-0.12}$ &  $20^{+20}_{-20}$ &  $51^{+11}_{-11}$ & $0.39^{+1.02}_{-1.02}$ &  -11.12 &   m & AO & LA09  \\
Q0449-BX93 &    $13^{+8}_{-8}$ &    $71^{+5}_{-5}$ &   $0.18^{+0.11}_{-0.11}$ &  $22^{+13}_{-13}$ &  $48^{+17}_{-17}$ &   $0.45^{+0.71}_{-0.71}$ &  -13.31 &   m & AO & LA09  \\
Q1217-BX95 &  $12^{+13}_{-13}$ &    $61^{+5}_{-5}$ &   $0.20^{+0.21}_{-0.21}$ &  $20^{+22}_{-22}$ &  $59^{+20}_{-20}$ &   $0.34^{+1.13}_{-1.13}$ &   10.17 &  dd & AO & LA09  \\
Q1623-BX376 &  $60^{+18}_{-18}$ & $99^{+11}_{-9}$ & $0.60^{+0.19}_{-0.19}$ & --- & --- & --- &  -12.03 &  rd & NS & FS09  \\
Q1623-BX447 & $100^{+30}_{-30}$ & $144^{+17}_{-17}$ & $0.69^{+0.22}_{-0.22}$ & --- & --- & --- &  -10.05 &  rd & NS & FS09  \\
Q1623-BX453 & $29^{+10}_{-10}$ & $94^{+5}_{-5}$ & $0.31^{+0.11}_{-0.11}$ &  $48^{+17}_{-17}$ & $66^{+20}_{-20}$ & $0.72^{+0.45}_{-0.45}$ & -0.16 & m & AO & LA09  \\
Q1623-BX455 &  $55^{+17}_{-17}$ & $130^{+28}_{-28}$ &   $0.42^{+0.16}_{-0.16}$ & --- & --- & --- &  -6.54 &  rd & NS & FS09  \\
\ditto & $125^{+15}_{-15}$ & $145^{+15}_{-16}$ & $0.86^{+0.14}_{-0.14}$ & $175^{+54}_{-17}$ & $56^{+15}_{-24}$ & $3.10^{+2.40}_{-0.50}$ & -6.54 &  rd & AO & FS18  \\
Q1623-BX502 &  $45^{+14}_{-14}$ & $72^{+7}_{-5}$ & $0.62^{+0.19}_{-0.20}$ & --- & --- & $0.80^{+0.20}_{-0.20}$ & 28.49 &  dd & NS & FS09  \\
\ditto & $35^{+11}_{-11}$ & $68^{+3}_{-3}$ & $0.51^{+0.16}_{-0.16}$ & $58^{+18}_{-18}$ & $57^{+17}_{-17}$ & $1.02^{+0.43}_{-0.43}$ & 28.49 &  dd & AO & LA09  \\
\ditto & $33^{+10}_{-10}$ & $66^{+5}_{-5}$ & $0.50^{+0.16}_{-0.16}$ &  $77^{+50}_{-32}$ & $40^{+16}_{-14}$ & $2.00^{+1.50}_{-0.80}$ & 28.49 &  rd & AO & FS18  \\
Q1623-BX528 & $67^{+20}_{-20}$ & $141^{+8}_{-8}$ & $0.48^{+0.14}_{-0.15}$ & --- & --- & --- & -11.53 &   m & NS & FS09  \\
Q1623-BX543 & $55^{+17}_{-17}$ & $149^{+22}_{-23}$ & $0.37^{+0.12}_{-0.12}$ & --- & --- & --- & -7.80 &   m & NS & FS09  \\
\ditto & $39^{+4}_{-4}$ & $153^{+7}_{-7}$ & $0.25^{+0.03}_{-0.03}$ & $65^{+7}_{-7}$ & $118^{+27}_{-27}$ & $0.55^{+0.25}_{-0.25}$ & -7.80 &   m & AO & LA09  \\
\ditto & $93^{+15}_{-15}$ & $163^{+8}_{-11}$ & $0.57^{+0.10}_{-0.10}$ & $128^{+29}_{-21}$ & $70^{+15}_{-25}$ & $1.80^{+0.90}_{-0.30}$ & -7.80 &   m & AO & FS18  \\
Q1623-BX599 & $49^{+15}_{-15}$ & $153^{+9}_{-9}$ & $0.32^{+0.10}_{-0.10}$ & --- & --- & --- & -0.66 &   m & NS & FS09  \\
\ditto & $80^{+20}_{-20}$ & $180^{+8}_{-10}$ & $0.44^{+0.11}_{-0.11}$ & $139^{+62}_{-36}$ & $71^{+18}_{-27}$ & $2.00^{+1.40}_{-0.40}$ & -0.66 &   m & AO & FS18  \\
Q1700-BX490 & $50^{+10}_{-10}$ & $122^{+6}_{-6}$ & $0.41^{+0.08}_{-0.08}$ &  $83^{+17}_{-17}$ &  $76^{+21}_{-21}$ & $1.09^{+0.34}_{-0.34}$ & 5.90 &   m & AO & LA09  \\
Q2343-BX389 & $205^{+62}_{-62}$ & $245^{+70}_{-50}$ & $0.84^{+0.30}_{-0.35}$ & --- & --- & $3.10^{+0.40}_{-0.40}$ & -8.30 &  rd & NS & FS09  \\
\ditto & $260^{+23}_{-23}$ & $258^{+23}_{-27}$ & $1.01^{+0.13}_{-0.14}$ & $299^{+40}_{-21}$ & $56^{+13}_{-15}$ & $5.30^{+1.70}_{-0.70}$ & -8.30 &  rd & AO & FS18  \\
Q2343-BX418 &  $23^{+12}_{-12}$ & $70^{+5}_{-5}$ & $0.33^{+0.17}_{-0.17}$ &  $38^{+20}_{-20}$ & $52^{+14}_{-14}$ & $0.73^{+0.59}_{-0.59}$ & 53.49 &  dd & AO & LA09  \\
Q2343-BX442 & $150^{+25}_{-25}$ & $180^{+20}_{-20}$ & $0.83^{+0.17}_{-0.17}$ & $234^{+49}_{-29}$ & $66^{+6}_{-6}$ & $3.55^{+0.23}_{-0.15}$ &  -15.00 &  rd & AO & LA12  \\
Q2343-BX513 & $27^{+8}_{-8}$ & $101^{+24}_{-19}$ & $0.26^{+0.09}_{-0.10}$ & --- & --- & --- & 18.43 &   m & NS & FS09  \\
\ditto &  $65^{+16}_{-16}$ & $102^{+6}_{-6}$ & $0.64^{+0.16}_{-0.16}$ & $108^{+26}_{-26}$ & $85^{+21}_{-21}$ & $1.27^{+0.35}_{-0.35}$ & 18.43 &   m & AO & LA09  \\
\ditto & $60^{+15}_{-15}$ & $139^{+11}_{-11}$ & $0.43^{+0.11}_{-0.11}$ & $102^{+64}_{-26}$ & $55^{+24}_{-28}$ & $1.80^{+2.10}_{-0.50}$ & 18.43 &   m & AO & FS18  \\
Q2343-BX660 & $40^{+9}_{-9}$ & $73^{+5}_{-5}$ & $0.55^{+0.13}_{-0.13}$ &  $66^{+15}_{-15}$ & $60^{+14}_{-14}$ & $1.10^{+0.32}_{-0.32}$ & 20.38 &   m & AO & LA09  \\
Q2346-BX404 &    $20^{+6}_{-6}$ & $97^{+4}_{-3}$ & $0.21^{+0.06}_{-0.06}$ & --- & --- & --- & 2.39 &  dd & NS & FS09  \\
Q2346-BX405 &  $32^{+10}_{-10}$ & $83^{+5}_{-3}$ & $0.39^{+0.12}_{-0.12}$ & --- & --- & --- &   -8.61 &  dd & NS & FS09  \\
Q2346-BX416 &  $70^{+21}_{-21}$ & $138^{+9}_{-9}$ & $0.51^{+0.16}_{-0.16}$ & --- & --- & --- & -17.22 &  rd & NS & FS09  \\
Q2346-BX482 & $233^{+70}_{-70}$ & $132^{+8}_{-9}$ & $1.76^{+0.54}_{-0.54}$ & $239^{+40}_{-40}$ & $52^{+13}_{-21}$ & $4.40^{+1.00}_{-1.00}$ &  -14.98 &  rd & NS & FS09  \\
\ditto & $225^{+20}_{-20}$ & $123^{+6}_{-7}$ & $1.83^{+0.19}_{-0.19}$ & $287^{+63}_{-30}$ &  $58^{+14}_{-15}$ & $4.90^{+1.60}_{-0.70}$ &  -14.98 &  rd & AO & FS18  \\
\bottomrule
\end{tabular}
\begin{tablenotes}
\item [a]  Half the observed difference between the maximum and minimum relative \halpha\ velocities across the source, uncorrected for inclination.  Equivalent to the shear velocity (${v}_{\mathrm{shear}}$) in the LA09 and LA12 nomenclature.
\item [b] Integrated velocity dispersion derived from the width of the \halpha\ emission line in the spatially collapsed object spectrum. Equivalent to ${\sigma }_{\mathrm{tot}}$ in FS18 and ${\sigma }_{\mathrm{net}}$ in LA09 and LA12.
\item [c] We use \vobs\ as an empirical measure of dynamical support and adopt the criterion of FS09 in which galaxies with \vobs $> 0.4$ are rotation-dominated, and those with \vobs $< 0.4$ are dispersion-dominated. 
\item [d] Intrinsic rotation velocity corrected for beam smearing and inclination angle according to ${v}_{\mathrm{rot}} \cdot \sin(i) = C_{\mathrm{PSF,v}} \cdot v_{\mathrm{obs}}/2$ (FS18).  For LA09 galaxies, we use $C_{\mathrm{PSF,v}} = 1.3$, the average correction factor of the FS18 sample, and $\sin(i) = \pi/4$, the mean inclination angle correction for a distribution of randomly inclined discs (LA09).
\item [e] Intrinsic local velocity dispersion determined from correction of the observed local velocity dispersion ($\sigma_{\mathrm{0obs}}$) for beam smearing according to $\sigma_{\mathrm{0}} = C_{\mathrm{PSF,}\sigma} \cdot \sigma_{\mathrm{0obs}}$ (FS18).  For the purposes of this comparison $\sigma_{\mathrm{mean}}$ in LA09 approximates to $\sigma_{\mathrm{0obs}}$ and we apply to the LA09 sample a correction of $C_{\mathrm{PSF},\sigma} = 0.85$, the average correction factor for the FS18 sample.
\item [f] Ratio of intrinsic rotation velocity and velocity dispersion corrected for beam smearing and inclination angle.  Within the rotating-disc framework of FS18, ${v}_{\mathrm{rot}}/{\sigma}_{\mathrm{0}} \approx \sqrt{3.36}$ corresponds to the point above which rotation starts to dominate over velocity dispersion in the dynamical support of turbulent discs.
\item [g] Kinematic classification assigned in the source studies (see Section~\ref{sec:c4_z2kin}):  rd = rotation-dominated, dd = dispersion-dominated, m = merger 
\item [h]  Observation mode employed in the source studies: NS = natural seeing, AO = adaptive optics assisted 
\item [i] Source references for kinematic data: FS09 = \citet{FS2009}, LA09 = \citet{Law2009}, LA12 = \citet{Law2012b}, FS18 = \citet{FS2018} 
\end{tablenotes} 
\end{threeparttable}}
\end{table*}

\begin{table*}
\centering
\caption{Kinematic and \lya\ Properties of UV Colour-Selected $z\sim3$ LBGs}
\label{tab:c4_table5}
\scalebox{0.85}{%
\begin{threeparttable}
\begin{tabular}{ccccccccccc}  
\toprule
\thead{ID} & 
\thead{${v}_{\mathrm{obs}}/2$\tnote{a} \\ (km\,s$^{-1}$)} & 
\thead{${\sigma}_{\mathrm{int}}$\tnote{b} \\ (km\,s$^{-1}$)} & 
\thead{${v}_{\mathrm{obs}}/2{\sigma}_{\mathrm{int}}$} & 
\thead{${v}_{\mathrm{rot}}$\tnote{c} \\ (km\,s$^{-1}$)} & 
\thead{${\sigma}_{\mathrm{0}}$\tnote{d} \\ (km\,s$^{-1}$)} & 
\thead{${v}_{\mathrm{rot}}/{\sigma}_{\mathrm{0}}$\tnote{e}} & 
\thead{net \lya\\ EW (\AA)} &  
\thead{Kin.\ \tnote{f}\\ class.} & 
\thead{Obs.\tnote{g} \\ mode} & 
\thead{Ref. \tnote{h}} \\
\midrule
SSA22a-D3 & $53\pm17$ & $115\pm13$ & $0.46\pm0.16$ & $109^{+52}_{-32}$ & $93^{+20}_{-36}$ & $1.16^{+0.72}_{-0.43}$ & 11.47 & rd & NS & TU17  \\
SSA22b-C20 & $66\pm18$ & $97\pm17$ & $0.68\pm0.22$ & $139^{+33}_{-30}$ & $48^{+34}_{-51}$ & $2.87^{+3.12}_{-2.14}$ & -4.18 & rd & NS & TU17  \\
SSA22b-D5 & $27\pm10$ & $38\pm20$ & $0.71\pm0.46$ & $57^{+26}_{-23}$ & $23^{+23}_{-18}$ & $2.45^{+3.44}_{-2.41}$ & 1.04 & rd & NS & TU17  \\
SSA22b-D9 & $32\pm14$ & $83\pm8$ & $0.39\pm0.17$ & $46^{+22}_{-15}$ & $73^{+10}_{-17}$ & $0.63^{+0.33}_{-0.23}$ & 37.83 & dd & NS & TU17  \\
SSA22b-MD25 & $44\pm15$ & $87\pm19$ & $0.51\pm0.20$ & $57^{+34}_{-24}$ & $75^{+21}_{-30}$ & $0.76^{+0.56}_{-0.39}$ & 32.81 & dd & NS & TU17  \\
SSA22a-M38 & --- & --- & --- & $346^{+51}_{-45}$ & $50^{+15}_{-15}$ & $6.92^{+2.31}_{-2.26}$ & -16.52 & rot & NS & GN11  \\
SSA22a-C16 & --- & --- & --- & $129^{+95}_{-24}$ & 90 & $1.43^{+1.06}_{-0.27}$ & -21.24 & rot & NS & GN11  \\
SSA22a-D17 & --- & --- & --- & $260^{+75}_{-65}$ & $130^{+16}_{-10}$ & $2.00^{+0.63}_{-0.52}$ & 18.14 & rot & NS & GN11  \\
CDFA-C9 & --- & --- & --- & $129^{+206}_{-31}$ & $117^{+3}_{-4}$ & $1.10^{+1.76}_{-0.27}$ & -6.48 & rot & NS & GN11  \\
3C324-C3 & --- & --- & --- & $86^{+63}_{-20}$ & $95^{+3}_{-5}$ & $0.91^{+0.66}_{-0.22}$ & 0.97 & rot & NS & GN11  \\
Q0302-C131 & --- & --- & --- & $>117$ & $43^{+5}_{-6}$ & $2.72^{+0.32}_{-0.38}$ & 27.60 & rot & AO & GN11  \\
\bottomrule
\end{tabular}
\begin{tablenotes}
\item [a] Observed rotation velocity measured directly from the 2D kinematic maps uncorrected for inclination (V$_{\mathrm{obs}}$ in TU17).  
\item [b] Observed velocity dispersion corrected for instrumental resolution ($\sigma_{\mathrm{obs}}$ in TU17().
\item [c] Intrinsic rotation velocity (V$_{\mathrm{max}}$ in GN11 and V$_{\mathrm{C}}$ in TU17).
\item [d] Intrinsic velocity dispersion ($\sigma_{\mathrm{int}}$ in both GN11 and TU17).  Listed $\sigma_{\mathrm{0}}$ values and uncertainties for GN11 galaxies are estimated from Fig.\ 20 in GN11.
\item [e] Ratio of intrinsic rotation velocity and velocity dispersion corrected for beam smearing and inclination angle.  TU17 classifies galaxies as rotation- or dispersion-dominated if ${v}_{\mathrm{rot}}/{\sigma}_{\mathrm{0}}$ is greater than or less than 1, respectively. 
\item [f] Kinematic classification assigned in the source studies (see Section~\ref{sec:c4_z3kin}):  rd = `rotation-dominated', rot = `rotating' and dd = `dispersion-dominated'. 
\item [g]  Observation mode employed in the source studies: NS = natural seeing, AO = adaptive optics assisted 
\item [h] Source references for kinematic data: TU17 = \citet{Turner2017} and GN11 = \citet{Gnerucci2011}
\end{tablenotes} 
\end{threeparttable}}
\end{table*}

\begin{figure*}
\centering
\scalebox{0.58}[0.58]{\includegraphics{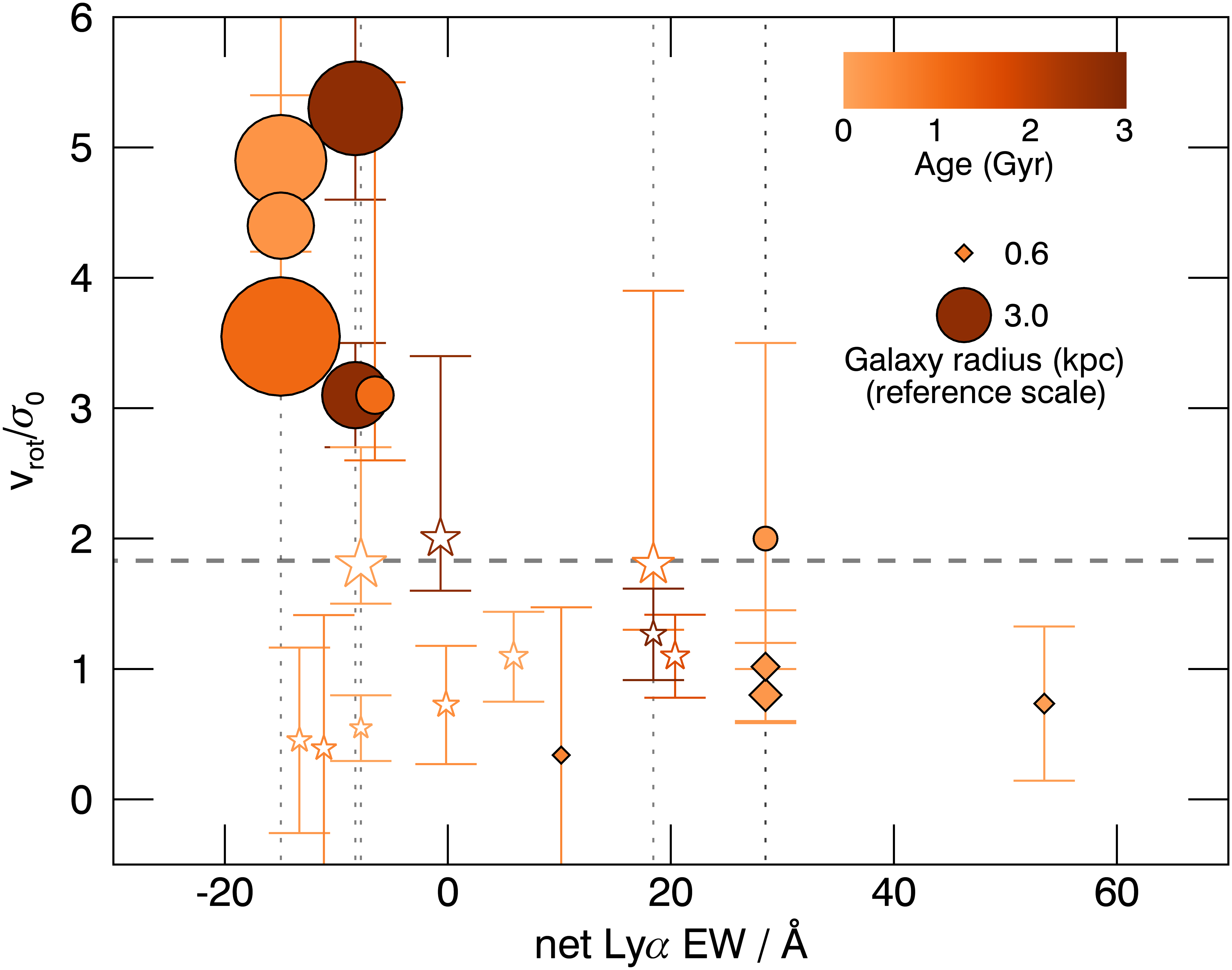}}
\scalebox{0.58}[0.58]{\includegraphics{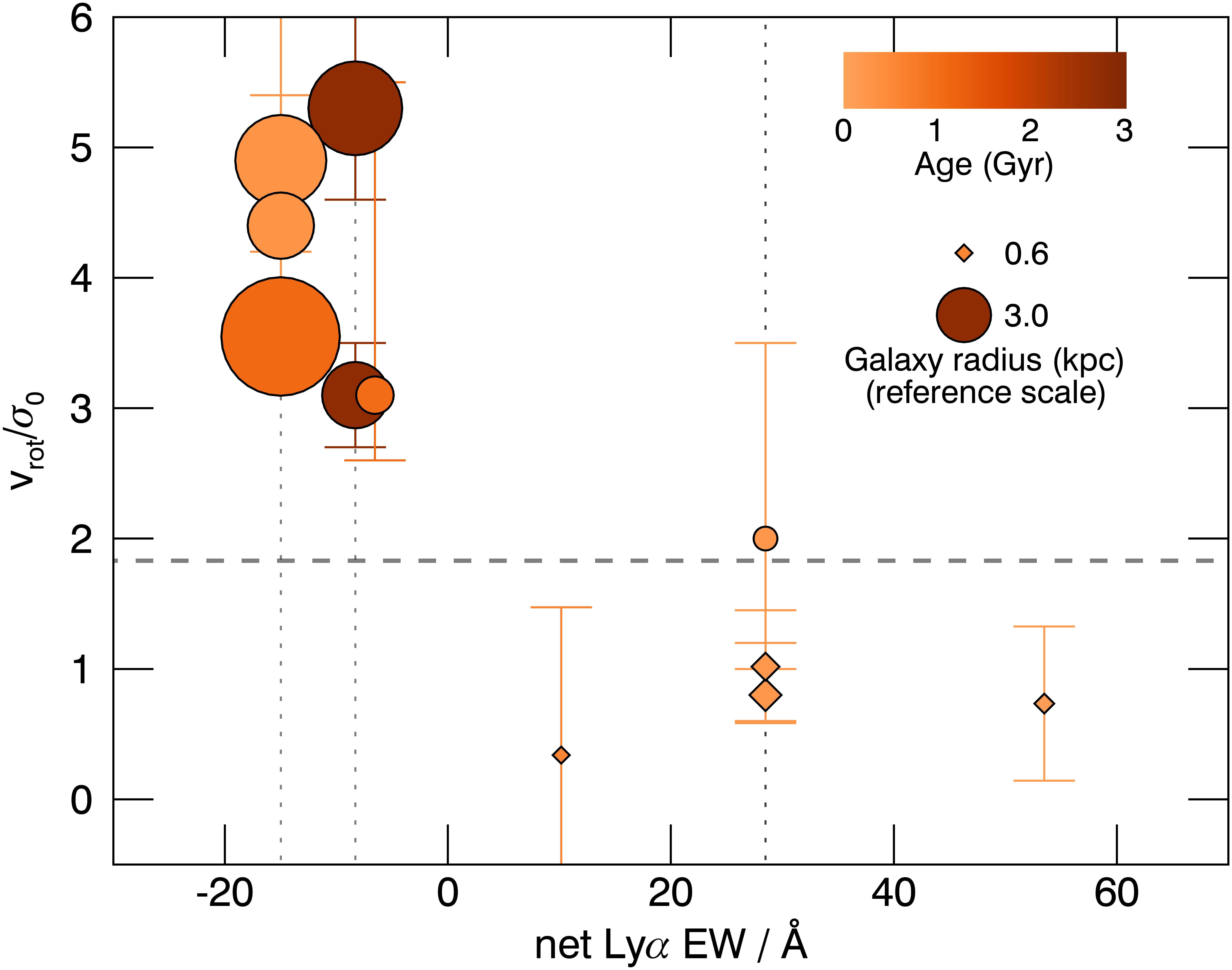}}
\scalebox{0.58}[0.58]{\includegraphics{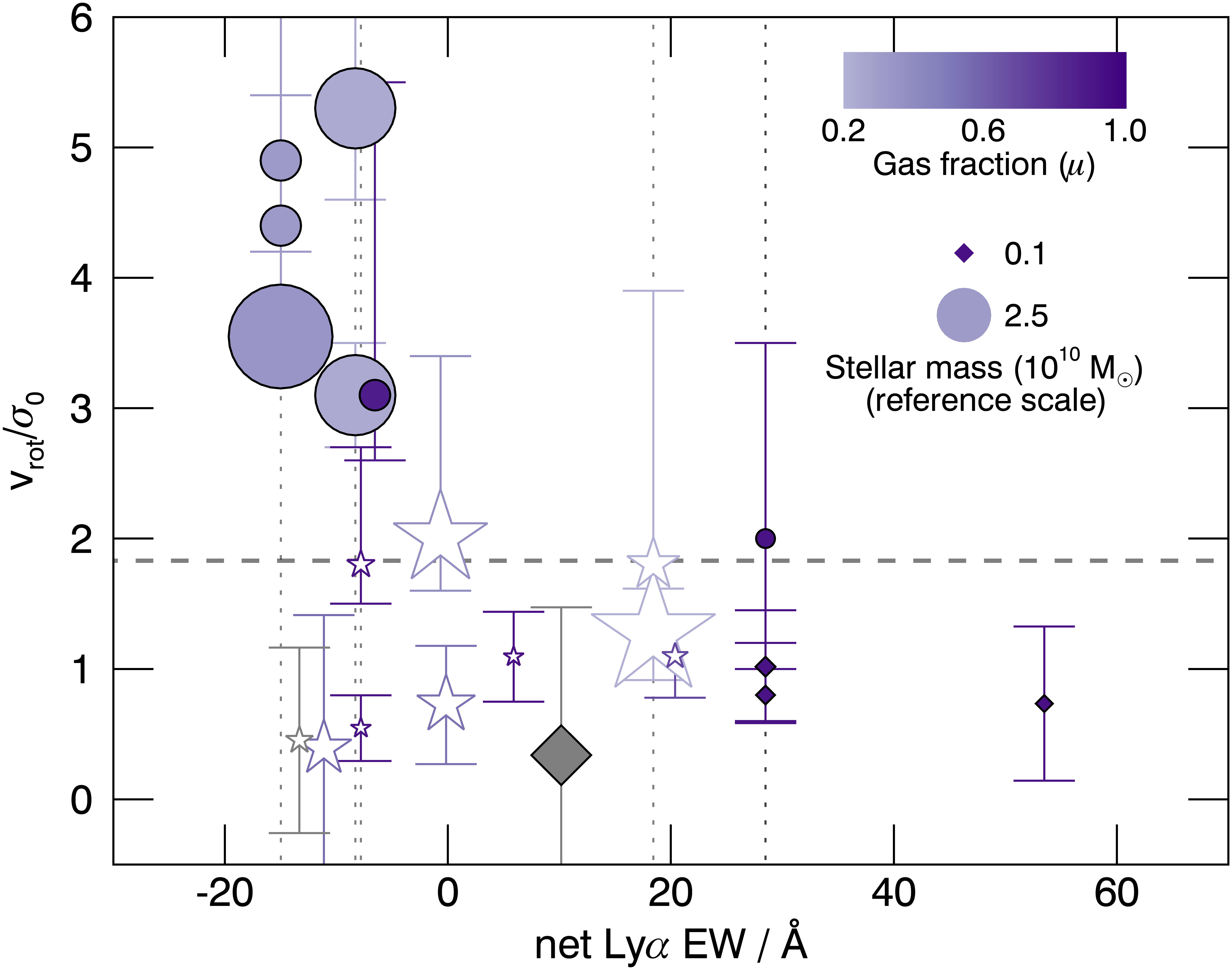}}
\scalebox{0.58}[0.58]{\includegraphics{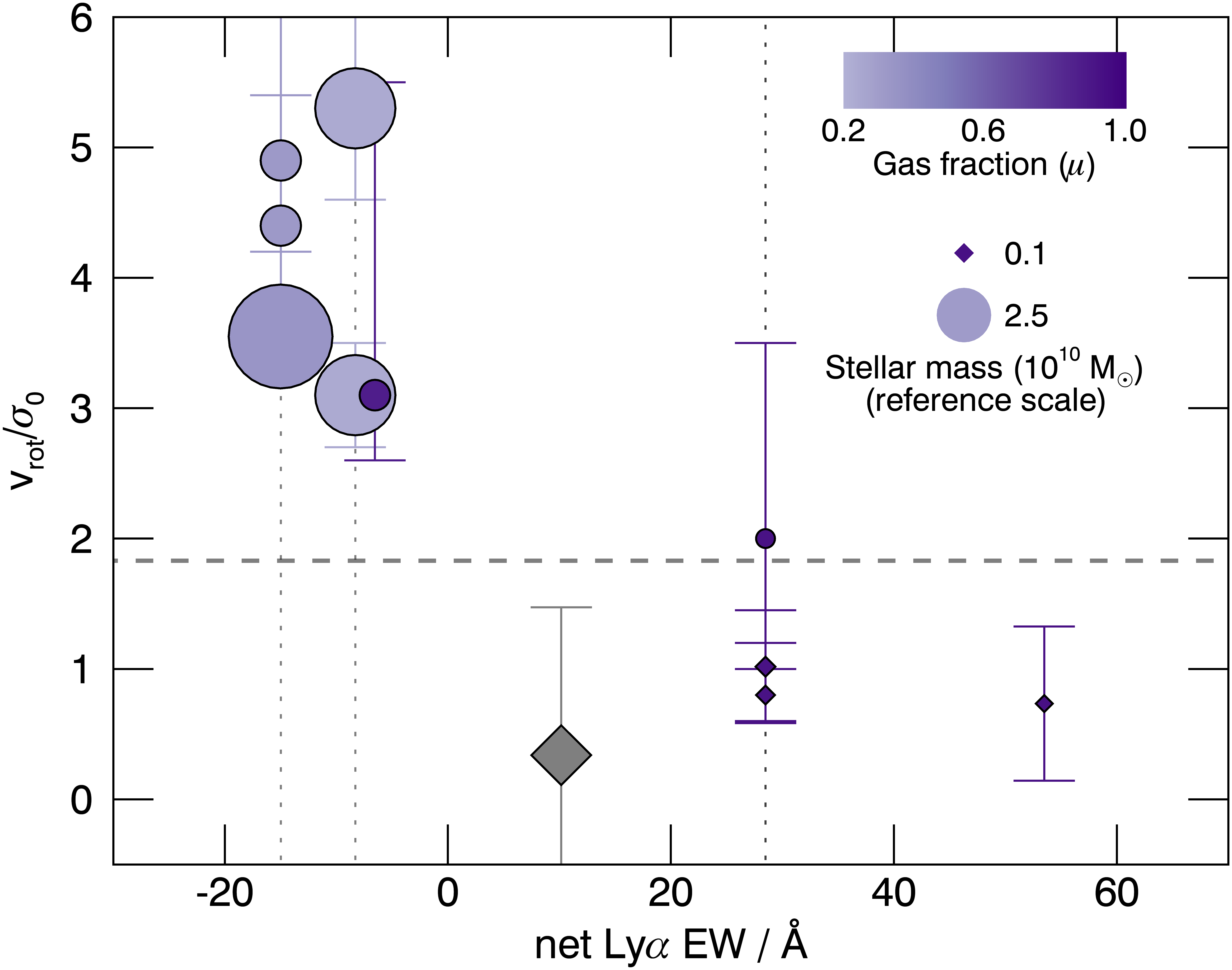}}
\scalebox{0.58}[0.58]{\includegraphics{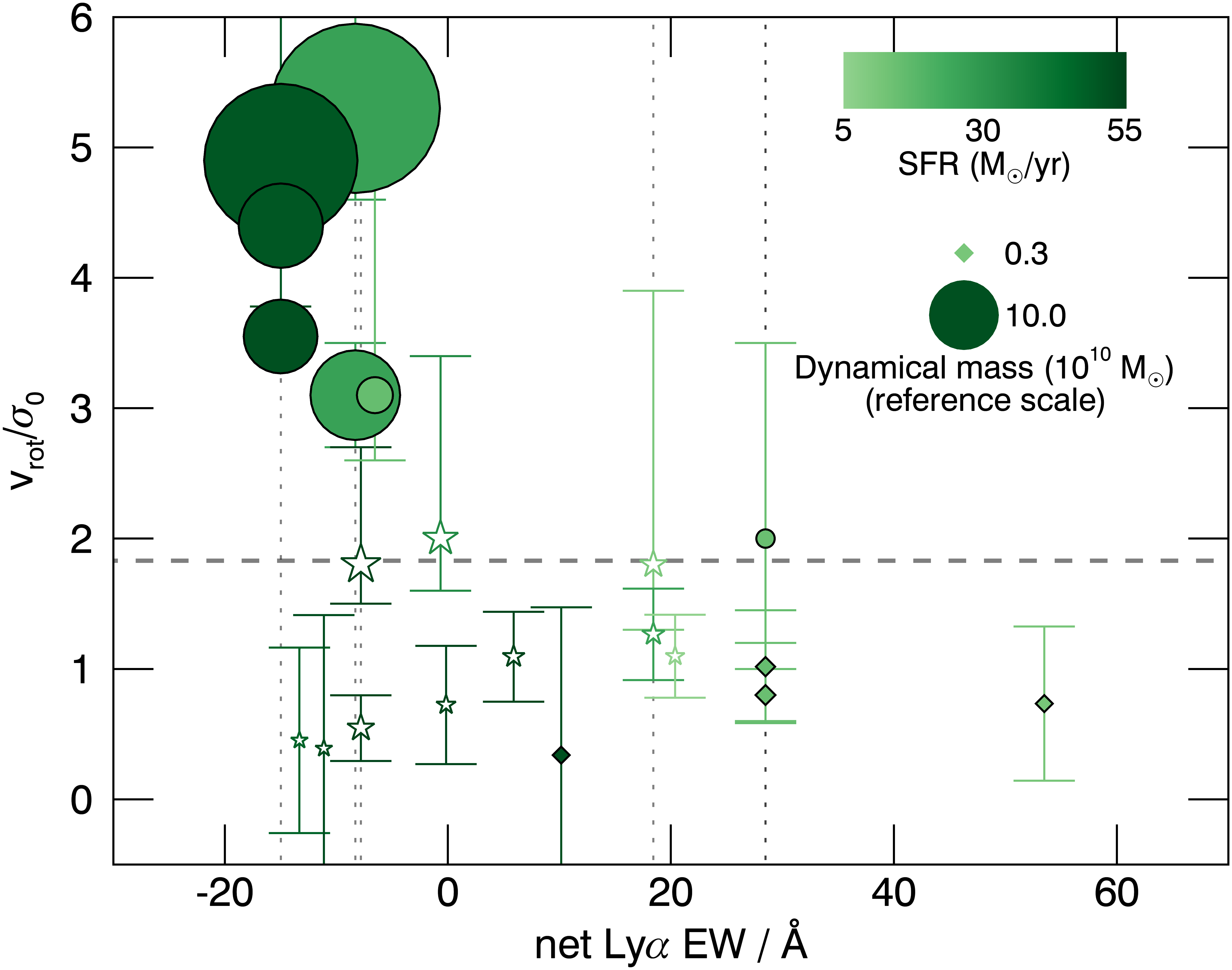}}
\scalebox{0.58}[0.58]{\includegraphics{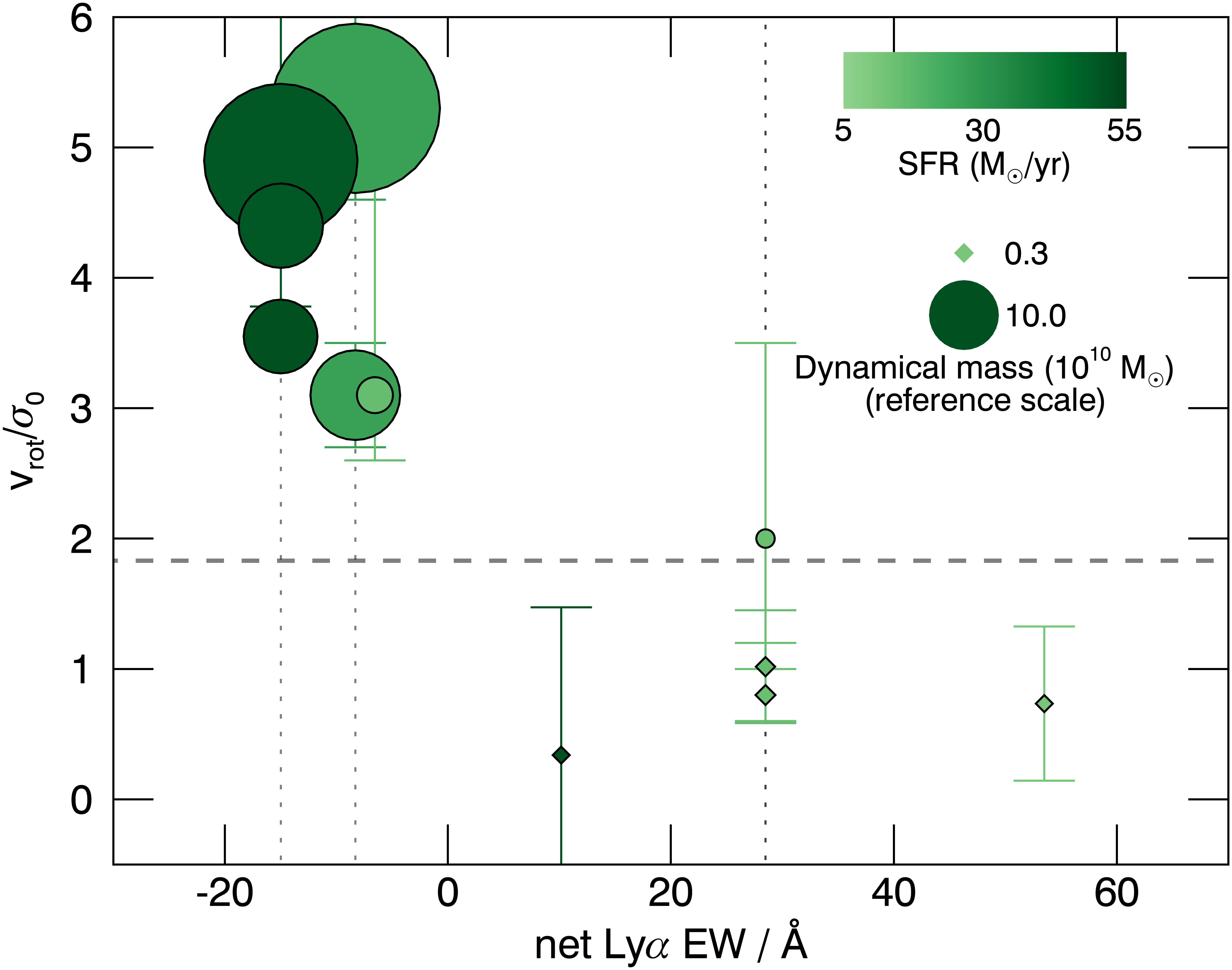}}
\caption{Physical properties of a sub-set of UV-colour-selected z $\sim2$ LBGs from FS09, LA09, LA12 and FS18 plotted as a function of the ratio of intrinsic rotation velocity to intrinsic velocity dispersion (\vrsO) and net \lya\ EW.  Kinematic classifications (symbol shapes), uncertainties, and broken lines are as described in Figure~\ref{fig:c4_fig6}.  Illustrative symbol sizes and colours encode the relative magnitude of galaxy physical properties (listed in Table~\ref{tab:c4_table4} and described in the text).  In each case, the plots show the full sample (left panels), as well as the sample with the mergers removed (right panels).  For what follows, the range of values rendered in each case for the non-merger sub-sample (filled symbols) are given in parentheses. Top: Galactic radius ($0.6-8.0$\,kpc) denoted by symbol size, and age from SED fitting ($0.10-2.75$\,Gyr) on a light to dark orange scale. Centre: Stellar mass ($0.1-6.0 \times 10^{10}$\,M$_{\odot}$) denoted by symbol size, and gas fraction ($0.22-0.93$) on a light to dark purple scale.  Grey symbols denote galaxies for which no value of $\mu$ is available. Bottom: Dynamical mass ($0.3-29 \times 10^{10}$\,M$_{\odot}$) denoted by symbol size, and star formation rate from SED fitting($11-52$\,M$_{\odot}$/yr) on a light to dark green scale.}
\label{fig:c4_fig7}
\end{figure*}

\subsubsection{Other galaxy properties}
\label{sec:c4_others}

In Figure~\ref{fig:c4_fig6}, we present net \lya\ EW as it relates to broadband colour and kinematics.  Figure~\ref{fig:c4_fig7} plots net \lya\ EW and kinematics in relation to several other properties previously reported for the galaxies in our sample: specifically, stellar mass ($M_{\star }$), star-formation rate from SED fitting (SFR$_{\mathrm{SED}}$), galactic size ($R_e$), age, gas fraction ($\mu$), and dynamical mass ($M_{dyn}$).  In each case, the plots show the full sample (left panels), as well as the sample with the mergers removed (right panels), in order to better understand the behaviour associated with discrete and isolated galaxies and those involved in interactions.  Symbol sizes are scaled as indicated in the caption and respective plot legends.  The range of colour gradients omits outliers to help highlight any systematic trends.  Table~\ref{tab:c4_table4} lists the kinematic, spectroscopic, and physical properties used to construct the plots.

In all cases, the quoted stellar properties ($M_{\star }$, SFR$_{\mathrm{SED}}$ and stellar population age) were derived from evolutionary synthesis modeling of optical to near-IR broadband spectral energy distributions (SEDs) supplemented with mid-IR photometry when available.  LA09,FS09 and FS18 used \citet{Bruzual2003} models with a \citet{Chabrier2003} initial mass function (IMF), the \citet{Calzetti2000} reddening law, and solar metallicity. Best fits were achieved with either constant or exponentially declining SFRs (see source studies for details).  The modelling procedures are described in detail by \citet{Shapley2005} (LA09) and in  Appendix A of FS09 (FS09, FS18).  For Q2343-BX442, LA12 used modelling procedures described by \citet{Erb2006b}, \citet{Shapley2005} and \citet{Reddy2010} to fit a SED constructed from ground-based $U_nG\mathcal{R}JK_s$, HST/WFC3 F160W and Spitzer IRAC photometry, with a Chabrier IMF, Calzetti reddening, and a constant star formation history.  LA12 used updated Charlot \& Bruzual population synthesis models that resulted in slightly lower estimates for $M_{\star }$, SFR$_{\mathrm{SED}}$ and stellar population age than estimates in the literature for similar samples \citep[cf.\ ][]{Erb2006b}.

For the SINS/zC-SINFONI galaxies of FS09 and FS18, we use the galactic half-light radius (R$_{e}$) values estimated from HST rest-frame optical (H-band) imaging  where available from \citet{Tacchella2015}.  Otherwise, sizes shown are \halpha\ radii derived from IFU maps.  In any event, the difference in sizes between \halpha\ and H-band emission is small, such that the choice made here is inconsequential (FS18).  For LA09 galaxies, R$_e$ is the radius of nebula emission.  For Q2343-BX442, we use the total luminous radius of $\sim8$\,kpc estimated by LA12 from HST/WFC3 infra-red imaging of the stellar continuum. 

To facilitate direct comparison across our sample, we use cold gas fractions ($\mu$ = $M_{gas}$/($M_{gas}$ + $M_{\star }$)) from \citet{Erb2006b} who estimated the mass of the gas associated with star formation using the local empirical correlation between star formation rate per unit area and gas surface density \citep{Kennicutt1998}.  Dynamical masses ($M_{dyn}$) for SINS/zC-SINFONI galaxies are derived from modelling in the `rotating-disc' framework of FS18, or by one of the related methods described in detail by FS09 for galaxies unsuitable for modelling as discs.  LA12 used a similar kinematic disc-fitting algorithm to model the velocity field of Q2343-BX442 and derive estimates for $M_{dyn}$.  For LA09 galaxies, $M_{dyn}$ is the single-component dynamical mass within the radius probed by nebular emission derived from the integrated velocity dispersion as per \citet{Erb2006b}.

The top panels show that rotation-dominated galaxies are larger and most are older than their dispersion-dominated counterparts.  There is no clear trend among the \lya-absorbing rotators, but the \lya-emitting dispersion-dominated galaxies are all similarly young.  The centre panels show that the more strongly-rotating \lya-absorbers typically have lower gas fractions and higher stellar masses than most of the dispersion-dominated \lya-emitters, but there is no apparent trend in either of these properties within the rotation-dominated sub-sample.  Although the dispersion-dominated galaxies have consistently high gas fractions and usually lower stellar masses, there is significant scatter in the stellar mass/net \lya\ EW trend in our small sample.  For example,galaxy Q1217-BX95 (large grey diamond in the centre panels) has relatively strong \lya\ emission (net \lya\ EW = $+$10.2\,\AA), dispersion-dominated kinematics (\vrsO\ = 0.34), but a relatively large stellar mass ($\sim3\times10^{10}$\,M$_{\odot}$) in the context of our sample. Finally, the bottom panels show larger dynamical masses and star formation rates for rotation-dominated sources compared to dispersion-dominated galaxies.

The small size of the kinematic sample that meets the necessary selection criteria for this work (see Section~\ref{sec:c4_overview}) prevents the robust investigation of the relationship between net \lya\ EW and kinematics at fixed values of other galactic properties, but even with this sample, we can see the emergence of trends that are consistent with relationships reported separately between kinematics, net \lya\ EW and the respective physical properties.  These illustrative plots suggest the potential value of a holistic multi-dimensional approach to the study of galaxy evolution in large samples and on large scales, and provide motivation to collect larger samples with consistent data that would inform the strength of these (and other) relations and how they are connected causally or otherwise.

\begin{table*}
\centering
\caption{Physical and Kinematic Properties of UV Colour-Selected $z\sim2$ LBGs}
\label{tab:c4_table4}
\scalebox{0.75}{%
\begin{threeparttable}
\begin{tabular}{cccccccccccc}  
\toprule
\thead{ID} & 
\thead{${v}_{\mathrm{rot}}/{\sigma}_{\mathrm{0}}$\tnote{a}} & 
\thead{net \lya\\ EW (\AA)} &  
\thead{$(U_n-\cal{R})$\\ Colour} & 
\thead{$R_e$\tnote{b} \\(kpc)} & 
\thead{Age\tnote{c} \\ (Myr)} & 
\thead{$M_{\star }$\tnote{c} \\ (${10}^{10}$\,M$_{\odot}$)} & 
\thead{$\mu$\tnote{d}} & 
\thead{$M_{dyn}$\tnote{e} \\ ($\times{10}^{10}$\,M$_{\odot}$)} & 
\thead{SFR$_{\mathrm{SED}}$\tnote{c} \\ (M$_{\odot}$\,yr$^{-1}$)} &  
\thead{Kin.\ \tnote{f}\\ class.} & 
\thead{Ref. \tnote{g}} \\
\midrule
HDF-BX1564 & $0.39^{+1.02}_{-1.02}$ & -11.12 & 1.28 & $1.3^{+0.2}_{-0.2}$ & $571^{+285}_{-285}$ & $2.90^{+1.16}_{-1.16}$ & 0.55 & 0.5 & $53^{+31}_{-31}$ & m & LA09  \\ 
Q0449-BX93 & $0.45^{+0.71}_{-0.71}$ & -13.31 & 0.63 & $1.3^{+0.1}_{-0.1}$ & $203^{+101}_{-101}$ & $0.90^{+0.36}_{-0.36}$ & --- & 0.5 & $47^{+28}_{-28}$ & m & LA09  \\
Q1217-BX95 & $0.34^{+1.13}_{-1.13}$ & 10.17 & 1.21 & $0.6^{+0.1}_{-0.1}$ & $571^{+285}_{-285}$ & $3.00^{+1.20}_{-1.20}$ & --- & 0.3 & $52^{+31}_{-31}$ & dd & LA09  \\
Q1623-BX453 & $0.72^{+0.45}_{-0.45}$ & -0.16 & 1.47 & $1.4^{+0.1}_{-0.1}$ & $404^{+202}_{-202}$ & $3.10^{+1.24}_{-1.24}$ & 0.52 & 1.0 & $77^{+46}_{-46}$ & m & LA09  \\
Q1623-BX455 & $3.1^{+2.4}_{-0.5}$ & -6.54 & 1.22 & $2.1^{+0.8}_{-0.4}$ & $1015^{+885}_{-611}$ & $1.03^{+0.52}_{-0.39}$ & 0.87 & $3.9^{+2.1}_{-0.7}$ & $15^{+10}_{-1}$ & rd & FS18  \\
Q1623-BX502 & $0.80^{+0.20}_{-0.20}$ & 28.49 & 0.72 & $1.7^{+0.5}_{-0.5}$ & $227^{+177}_{-137}$ & $0.23^{+0.14}_{-0.10}$ & 0.92 & $0.85^{+0.36}_{-0.36}$ & $14^{+9}_{-1}$ & dd & FS09  \\
\ditto & $1.02^{+0.43}_{-0.43}$ & 28.49 & 0.72 & $1.4^{+0.2}_{-0.2}$ & $203^{+101}_{-101}$ & $0.30^{+0.12}_{-0.12}$ & 0.92 & 0.7 & $15^{+9}_{-9}$ & dd & LA09  \\
\ditto & $2.0^{+1.5}_{-0.8}$ & 28.49 & 0.72 & $1.1^{+0.7}_{-0.7}$ & $227^{+177}_{-137}$ & $0.23^{+0.14}_{-0.10}$ & 0.92 & $0.58^{+0.77}_{-0.21}$ & $14^{+9}_{-1}$ & rd & FS18  \\
Q1623-BX543 & $0.55^{+0.25}_{-0.25}$ & -7.80 & 1.40 & $1.1^{+0.1}_{-0.1}$ & $10^{+5}_{-5}$ & $0.40^{+0.16}_{-0.16}$ & 0.92 & 2.5 & $431^{+258}_{-258}$ & m & LA09  \\
\ditto & $1.8^{+0.9}_{-0.3}$ & -7.80 & 1.40 & $3.3^{+1.2}_{-0.8}$ & $81^{+21}_{-9}$ & $0.94^{+0.20}_{-0.09}$ & 0.92 & $5.0^{+2.4}_{-2.0}$ & $150^{+2}_{-5}$ & m & FS18  \\
Q1623-BX599 & $2.0^{+1.4}_{-0.4}$ & -0.66 & 1.02 & $2.4^{+0.6}_{-0.6}$ & $2750^{+18}_{-1141}$ & $5.66^{+0.07}_{-0.24}$ & 0.37 & $4.1^{+3.3}_{-1.7}$ & $34^{+19}_{-1}$ & m & FS18  \\
Q1700-BX490 & $1.09^{+0.34}_{-0.34}$ & 5.90 & 1.28 & $1.6^{+0.1}_{-0.1}$ & $10^{+5}_{-5}$ & $0.40^{+0.16}_{-0.16}$ & 0.93 & 1.5 & $441^{+264}_{-264}$ & m & LA09  \\
Q2343-BX389 & $3.1^{+0.4}_{-0.4}$ & -8.30 & 1.54 & $4.2^{+1.1}_{-1.1}$ & $2750^{+224}_{-250}$ & $4.40^{+0.19}_{-0.31}$ & 0.22 & $14^{+1}_{-1}$ & $26^{+1}_{-2}$ & rd & FS09  \\
\ditto & $5.3^{+1.7}_{-0.7}$ & -8.30 & 1.54 & $6.2^{+1.2}_{-1.2}$ & $2750^{+224}_{-250}$ & $4.40^{+0.19}_{-0.31}$ & 0.22 & $29^{+9}_{-6}$ & $26^{+1}_{-2}$ & rd & FS18  \\
Q2343-BX418 & $0.73^{+0.59}_{-0.59}$ & 53.49 & 0.32 & $0.8^{+0.1}_{-0.1}$ & $102^{+51}_{-51}$ & $0.10^{+0.04}_{-0.04}$ & 0.93 & 0.3 & $11^{+6}_{-6}$ & dd & LA09  \\
Q2343-BX442 & $3.55^{+0.23}_{-0.15}$ & -15.00 & 1.54 & $\sim8$ & $1100^{+1000}_{-500}$ & $6^{+2}_{-1}$ & 0.33 & 11 & $52^{+37}_{-21}$ & rd & LA12  \\
Q2343-BX513 & $1.27^{+0.35}_{-0.35}$ & 18.43 & 0.61 & $1.4^{+0.1}_{-0.1}$ & $3000^{+1500}_{-1500}$ & $7.80^{+3.12}_{-3.12}$ & 0.19 & 1.6 & $26^{+15}_{-15}$ & m & LA09  \\
\ditto & $1.8^{+2.1}_{-0.5}$ & 18.43 & 0.61 & $2.6^{+1.5}_{-0.8}$ & $806^{+209}_{-166}$ & $2.70^{+0.56}_{-0.05}$ & 0.19 & $2.5^{+3.2}_{-1.2}$ & $10^{+1}_{-4}$ & m & FS18  \\
Q2343-BX660 & $1.10^{+0.32}_{-0.32}$ & 20.38 & 0.36 & $1.6^{+0.1}_{-0.1}$ & $1609^{+804}_{-804}$ & $0.80^{+0.32}_{-0.32}$ & 0.69 & 0.9 & $5^{+3}_{-3}$ & m & LA09  \\
Q2346-BX482 & $4.4^{+1.0}_{-1.0}$ & -14.98 & 1.12 & $4.2^{+0.9}_{-0.9}$ & $286^{+35}_{-83}$ & $1.69^{+0.30}_{-0.25}$ & 0.31 & $13^{+2}_{-1}$ & $50^{+32}_{-1}$ & rd & FS09  \\
\ditto & $4.9^{+1.6}_{-0.7}$ & -14.98 & 1.12 & $6.0^{+0.8}_{-0.8}$ & $286^{+35}_{-83}$ & $1.69^{+0.30}_{-0.25}$ & 0.31 & $26^{+12}_{-5}$ & $50^{+32}_{-1}$ & rd & FS18  \\
\bottomrule
\end{tabular}
\begin{tablenotes}
\item [a]  Ratio of intrinsic rotation velocity to intrinsic velocity dispersion corrected for beam smearing and inclination angle (see Section~\ref{sec:c4_kinemetry})  
\item [b]  Galactic half-light radius  
\item [c]  Stellar properties from SED modelling (mass ($m_{\star }$), age, and star-formation rate (SFR$_{\mathrm{SED}}$)) collated from the respective source studies  
\item [d]  Cold gas fraction ($\mu$ = $M_{gas}$/($M_{gas}$ + $M_{\star}$)) from \citet{Erb2006c}  
\item [e]  Dynamical mass
\item [f]  Kinematic classification assigned in the source studies: rd = rotation-dominated, dd = dispersion-dominated, m = merger 
\item [g]  Source references for kinematic and other data: FS09 = \citet{FS2009}, LA09 = \citet{Law2009}, LA12 = \citet{Law2012b}, FS18 = \citet{FS2018} 
\end{tablenotes} 
\end{threeparttable}}
\end{table*}

\section{DISCUSSION} 

\subsection{Context}

The results presented here are part of a broader project that aims to examine the intrinsic and environmental properties of high redshift galaxies in a holistic manner to lend insight into their evolution.  C09 and Paper\,I describe a method by which large samples of $z\sim2-3$ LBGs with known \lya\ spectral type can be selected from optical broadband imaging datasets, and by so doing suggest a means by which   a wide range of other spectral, physical, and environmental galaxy properties known to correlate with \lya\ might be studied in large samples and on large scales using optical broadband imaging and/or \lya\ data alone.  The power of this approach was demonstrated by \citet{Cooke2013} who studied the large-scale clustering properties and halo mass distributions of photometrically selected $z\sim3$ aLBG and eLBG populations via correlation function analysis (see Section~\ref{sec:c4_lss}).

\lya-sensitive properties accessible by this approach include: size and morphology \citep[e.g.,][]{Law2007a, Law2012c, Pentericci2010, Marchi2019}; gas fraction \citep[e.g., ][]{Erb2006b, Law2012c}; parameters derived from SED fitting such as stellar mass, SFR, UV continuum slope/reddening and age \citep[e.g.,][]{Shapley2003, Erb2006a, Law2007a, Kornei2010, Steidel2010, Law2012c, Hathi2016, Reddy2016, Du2018, Marchi2019}; gas covering fraction and outflow kinematics \citep[e.g., ][]{Shapley2003, Steidel2010, Law2012c, Reddy2016, Guaita2017, Steidel2018, Du2018, Trainor2019, Marchi2019}; low and high ionisation absorption line and nebula emission line strengths \citep[e.g.,][]{Shapley2003, Steidel2010, Law2012c, Stark2014, Trainor2016, Erb2016, Guaita2017, Du2018}; and, dark matter halo mass, and large-scale spatial distribution \citep[e.g.,][]{Cooke2013, Diaz2014, Guaita2017, Guaita2020, Shi2019}.

In this work we add the relationships between \lya, broadband colour, and nebular emission-line kinematics to this list.  

\subsection{The importance of kinematics}

Galaxy kinematics are a characteristic feature of the Morphology--Density Relation (MDR) in the local Universe \citep[][and references therein]{Bamford2009}, and are a key constraint for simulations that aim to understand the mechanisms by which galaxies evolve over cosmic time \citep{Somerville2015}.  Despite its critical role in reconciling observations with physical and computational models, acquiring kinematic information for thousands to millions of galaxies over large scales from high to low redshift is not possible for the foreseeable future -- in particular for galaxies at $z\gtrsim 4$.  Thus, there is strong motivation to explore connections between kinematics and other properties that may serve as proxies for predicting kinematic type where such detail is not directly obtainable.

At $z \gtrsim 4$, the bulk of the galaxy population is only accessible via deep imaging and, for galaxies reachable with deep spectroscopy, often \lya\ is the only accessible feature at optical/near-IR wavelengths.  This is also the case for the fainter and lower-mass $z\sim2-3$ galaxies.  In this work we have reported a fundamental relationship between galaxy kinematics and net \lya\ EW that has potential to serve as a predictor of kinematic type in such cases.  In addition, we have proposed a broadband imaging method whereby samples of known kinematic type can be selected via the photometric segregation of \lya\ spectral types on the CMD, and from which the most promising galaxies can be efficiently selected for expensive follow-up observations.  We do not propose that the coarse kinematic behaviour derived from this method is a substitute for the detailed and more accurate kinematic information available from IFU-based studies.  Rather, we suggest this method as a complementary approach that can elucidate the typical kinematic character of large samples for which individual IFU measurements are not feasible, and at redshifts $z \gtrsim 4$ where such measurements are currently not possible.

Together, these results provide a means to explore all the above \lya-sensitive properties, and their relation to generalised kinematic type in very large samples of galaxies on large scales and potentially out to high redshift in datasets from current and future large-area photometric campaigns.  For example, we envisage application of our method to datasets from the all-sky LSST that will select hundreds of millions of LBGs in redshift ranges from $z\sim2-6$ across many hundreds to thousands of Mpc. 

\subsection{Comparison with low-redshift analogues}
\label{sec:c4_lowz}

The fourteen low redshift ($0.03<z<0.2$) star-forming galaxies comprising the \lymana\ Reference Sample (LARS) have continuum size, stellar mass, and rest-frame absolute magnitudes typical of Lyman break analogues in the local Universe selected to have properties similar to star-forming galaxies at $2<z<3$ \citep{Ostlin2014, Hayes2014}.  Accordingly, they provide a useful low-z reference sample for comparison with our $z\sim2-3$ results.  

\citet[][hereafter HE16]{Herenz2016} derive values for shear velocity (v$_{\rm{shear}}$) and intrinsic velocity dispersion ($\sigma_0$) from the \halpha\ kinematic maps of the LARS galaxies in a manner similar to the methods used by FS09, LA09 and FS18 at $z\sim2$.  They classify each galaxy as either a `rotating disc', `perturbed rotator', or as having `complex kinematics' based on the qualitative appearance of their velocity fields and the classification scheme introduced by \citet{Flores2006}.  HE16 show that the LARS galaxies are characterised by high intrinsic velocity dispersions in the range 40$-$100\kms\ (54\kms\ median), low shear velocities (30$-$180\kms, 65\kms\ median), and v$_{\rm{shear}}/\sigma_0$ ratios ranging from 0.5 to 3.2.  In this respect the LARS galaxies are kinematically similar to turbulent star-forming galaxies observed at high redshift (HE16), including the $z\sim2-3$ LBGs studied herein (cf.\ the kinematic parameters in Tables~\ref{tab:c4_table3}~\&~\ref{tab:c4_table5}). 

The LARS team used a synthetic narrow-band imaging method to measure total flux in the region of \lya, and the subtraction from this of a modelled stellar continuum spectrum to derive the \lya\ flux (or flux decrement).  Pixel-wise SED fitting of the same spectral models as were used for continuum subtraction was then used to estimate the continuum flux density at 1216\,\AA\ by which the flux differential was divided in order to calculate a \lya\ EW.  The bandpass used to measure \lya\ sampled rest-frame wavelengths between about 1205 and 1230\,\AA\ (depending on the exact redshift of the target galaxy).  Accordingly, the measured flux included both \lya\ in emission and at least part of the \lya\ absorption signal blueward of 1216\,\AA.  For comparison with our $z\sim2-3$ samples, we use the integrated values of \lya\ EW given by HE16 and \citet{Hayes2014} that were calculated using fluxes and flux densities integrated over a circular aperture with twice the isophotal Petrosian radius determined for each galaxy from the image that transmits \lya\ and the far-UV continuum (see \citet{Hayes2013} and \citet{Hayes2014}).

LARS galaxies with higher shearing velocities (v$_{\rm{shear}} \gtrsim 50$\kms) have preferentially lower \lya\ EWs and lower \lya\ escape fractions than their lower angular momentum counterparts.  Moreover, \lya\ EW and \lya\ escape fraction correlate with v$_{\rm{shear}}/\sigma_0$ in the sense that the LARS galaxies with `complex kinematics' have higher \lya\ EWs and higher \lya\ escape fractions than systems with a kinematic signature indicative of a `perturbed rotator' or a `rotating disc'.  

While these observations of HE16 are in good qualitative agreement with our findings, the confidence with which we can directly compare the LARS \lya\ EW values with the net \lya\ EWs of the $z\sim2-3$ LBGs is moderated by a number of factors:

\begin{enumerate}[label=(\roman*)]
\item Due to the relatively broad bandpass of the synthetic filters used for the LARS imaging, a large amount of stellar continuum light is transmitted along with the \lya\ signal.  The reliability of the measured \lya\ flux is, therefore, critically dependent on the quality of the stellar continuum modelling, and in cases where stellar light dominates over \lya, small errors in the modelling can result in large errors in the derived properties (M.\ Hayes, priv.\ comm.).
\item \citet{Hayes2014} found that due to the spatial redistribution of \lya\ under the influence of resonant scattering processes, the computation of integrated \lya\ quantities for their galaxies is a strong function of the radius over which values were summed.  Figure 4 in \citet{Hayes2014} shows the aperture-dependent behaviour of \lya\ EW (and other properties) for the LARS galaxies.
\item The LARS team chose to adopt a convention of reporting \lya\ EW in emission only such that even if an integrated \lya\ EW with net absorption was measured, it has been reported only in terms of its emission component, i.e., with a \lya\ EW of zero (M.\ Hayes, priv.\ comm.).  
\end{enumerate}

Figure~\ref{fig:c4_fig6c} shows a v$_{\rm{shear}}/\sigma_0$ versus \lya\ EW plot of the HE16 galaxies overlaid on a \vrsO\ versus net \lya\ EW plot of our combined $z\sim2$ and $z\sim3$ kinematic samples.  The curve-of-growth results of \citet{Hayes2014} indicate that Two LARS galaxies (LARS04 \& LARS06) are net absorbers at all apertures within their field of view, but these are reported as having \lya\ EW = 0 under the convention adopted by the LARS team.  While the magnitude of any offset is difficult to quantify, these two galaxies would move toward the left on the plot if this net absorbing character was reflected in the quoted \lya\ EWs.  Taking into account this caveat, and the other sources of systematic uncertainty notwithstanding, the relationship between galaxy kinematics and \lya\ observables in the LARS sample is almost indistinguishable from our result at $z\sim2-3$.

HE16 surmise that there is a causal relationship between turbulence in actively star-forming galaxies and ISM conditions that facilitate the escape of \lya\ photons, and further speculate that dispersion-dominated kinematics are a necessary requirement for a galaxy to have a significant amount of escaping \lya\ radiation.  HE16 also note that, like the $z\sim2$ samples of FS09 and LA09, galaxies in their sample with lower stellar mass ($M_{\star }$) typically have lower v$_{\rm{shear}}/\sigma_0$ ratios and that the strongly \lya-emitting LARS-LAEs are preferentially found among the systems with $M_{\star }$ $\lesssim 10^{10}\,{M}_{\odot}$. 

These conclusions for the low-z LARS galaxies are consistent with our results that show a link between higher net \lya\ EW and dispersion-dominated kinematics in $z\sim2-3$ LBGs.  Moreover, they support our general proposition that \lya\ emission may be a useful diagnostic of galaxy kinematics and other properties, particularly at high redshifts where the number density of LAEs is higher \citep[e.g.,][]{Wold2014}, and low-mass star-forming dispersion-dominated galaxies are more prevalent \citep[e.g.,][]{Turner2017, Mason2017, Girard2018}.

\subsection{Implications for galaxy evolution science}
\label{sec:c4_implications}

\subsubsection{Kinematics and Large-Scale Structure at $z\sim2-3$}
\label{sec:c4_lss}

A key finding of this work, Paper\,I, and C09 is that \lya\ spectral types in populations of $z\sim2-3$ LBGs segregate consistently with a range of intrinsic galactic properties.  For example, we show in Sections~\ref{sec:c4_elbgs}~\&~\ref{sec:c4_others} that eLBGs are characteristically, compact, blue, low-mass, low-metallicity, presumably young, dispersion-dominated systems, and that aLBGs are typically, red, spatially-diffuse, high-mass rotation-dominated galaxies, usually with disc-like morphology.

\citet{Cooke2013} use the same photometric selection method to isolate large samples ($\sim$10$^5$) of $z\sim3$ aLBGs and eLBGs, and investigated their respective dark matter halo mass and spatial distribution on large scales using two-point correlation function analysis.  They find that aLBGs preferentially reside in group and cluster environments, and eLBGs are typically found on the outskirts of groups and clusters and in the field.  Moreover, cross-correlation function results showed that the two spectral types avoid each other on single halo to cluster halo scales.  Similarly, and consistent with simulations that predict a decrease in \lya\ EW with increasing overdensity in $z\sim2$ protoclusters \citep{Muldrew2015}, spectroscopic follow-up of protoclusters identified at $z\sim3-4$ show that protocluster members have lower average \lya\ EW compared to equivalent coeval field galaxies \citep{Toshikawa2016,Lemaux2018}.  There is also a growing body of work indicating that galaxies with large net \lya\ EW are over-represented in under-dense regions, while galaxies with lower net \lya\ EW are mainly located in over-dense environments \citep[e.g.,][]{Ouchi2010, Diaz2014, Bielby2016, Guaita2017, Guaita2020, Shi2019}. 

Combining these results with our findings, it would appear that we are seeing at $z\sim2-3$, a spatial segregation of galaxies with rotation- and dispersion-dominated kinematics on large ($\sim$100\,Mpc) scales.  That is, larger more massive rotating disc galaxies (aLBGs in our system) are located preferentially in group environments, and dispersion-dominated eLBGs tend to be found on group outskirts and in the field.  

\subsubsection{Bimodality at high redshift and the Morphology--Density Relation} 
\label{sec:c4_power}

The demonstrated segregation of \lya\ spectral types with a wide range of galactic properties -- including the large-scale clustering behaviour of aLBGs and eLBGs described above -- is suggestive of a non-homogeneous galaxy population at high redshift.  Indeed, it is reminiscent of the bimodal distribution of blue, star-forming spirals and large, massive red and quiescent ellipsoids of the modern-day universe (see for example \citet{Blanton2009}).

Independent evidence of galaxy bimodality at high redshift is observed in damped \lya\ systems (DLAs).  \citet{Wolfe2008} found a bimodality in the DLA population based on [C{\footnotesize II}] 158~$\rm \mu m$ cooling rates that is independent of H{\footnotesize I} column density.  DLAs with high cooling rates have significantly higher velocity line profiles, metallicity, dust to gas ratios, ISM line widths, and star formation rates compared to those with low cooling rates.  Intriguingly, the properties of the high and low cooling-rate DLAs map well with the properties of aLBG and eLBG \lya\ spectral types respectively, as discussed in this work and by \citet{Shapley2003}, \citet{Du2018}, and \citet{Pahl2020}.

These results caution against assuming that high-redshift galaxy populations are homogeneous, and emphasise the need to consider heterogeneity, and likely bimodality, in the distribution of galactic properties.  Accordingly, from a holistic consideration of the many relationships between \lya\ and galaxy properties discussed above, we envisage a model in which massive rotating disc systems (aLBGs) that are preferentially found in groups and clusters merge to form present-day elliptical galaxies \citep[e.g.,][]{Toomre1972}.  Compact eLBGs may be either super-star clusters in faint, low mass galaxies, or the precursors of bulges in present-day spiral galaxies, which are known to have dispersion-supported kinematics and are expected to form at high redshift.  Such a model suggests that we are observing at $z\sim2-3$ signs of a nascent morphology-density relation \citep{Dressler1980, Postman2005} traced by the aLBG and eLBG populations.

\section{Summary and Conclusions}

In this paper we report a direct relationship between nebular emission-line kinematics and net \lya\ EW in  samples of $z\sim2$ and $z\sim3$ LBGs drawn from the literature for which matching rest-frame UV broadband photometry, consistently measured net \lya\ EWs, and kinematic classifications from IFU-based spectroscopy are available.  We conclude that LBGs with \lya\ dominant in absorption (aLBGs) are almost exclusively rotation-dominated (presumably disc-like) systems, and LBGs with \lya\ dominant in emission (eLBGs) characteristically have dispersion-dominated kinematics.  

The key results of this paper are summarised below:

\begin{enumerate}[label=(\roman*)]
\item In Sections~\ref{sec:c4_z3lbgs}~\&~\ref{sec:c4_z2lbgs} we show that rotation- and dispersion-dominated $z\sim2-3$ LBGs segregate consistently with rest-frame UV colour, and that their distributions on a rest-frame UV colour-magnitude diagram (CMD) are coincident with the aLBG and eLBG distributions respectively of the parent LBG samples from which they were drawn.  
\item This congruent behaviour is reinforced by the assignment to the kinematic samples of \lya\ spectral types based on spectroscopically-determined net \lya\ EWs (see Section~\ref{sec:c4_lya_type}), and is statistically supported by the results of two-sided Kolmogorov--Smirnov (KS) tests. 
\item Galaxies located in the strongly \lya-absorbing part of the CMD (aLBGs) are characteristically massive, red, spatially diffuse disc-like systems.  Galaxies with photometric properties akin to those with \lya\ dominant in emission (eLBGs) are most likely to be compact, blue, and dispersion-dominated, with low (if any) rotational dynamic support, free from large or luminous rotating disc structures.  Moreover, we find that sources in our kinematic samples that have been positively identified as merging systems reside toward the bright side, and centrally in colour, on the CMD.
\item In Section~\ref{sec:c4_class} we report the segregation in average net \lya\ EW for subsets of rotation- and dispersion-dominated galaxies in our $z\sim2$ and $z\sim3$ kinematic samples, and show, for a combined $z\sim2-3$ sample of 32 galaxies, a clear bifurcation (KS-test confidence $\sim$99\%) in the average \lya\ spectral properties of rotation- and dispersion-dominated LBGs.
\item In Section~\ref{sec:c4_kinemetry} we quantify the  relationship between the strength of rotational dynamic support (as measured using \vobs\ and \vrsO) and net \lya\ EW for subsets of our kinematic sample where these data are available.   Both results show a statistically significant non-linear negative correlation between rotational dynamic support and net \lya\ EW.
\item In Section~\ref{sec:c4_others} we confirm that the relationship between net \lya\ EW and kinematics that we report is consistent with relationships reported separately between kinematics, net \lya\ EW and other galactic properties such as stellar mass, star-formation rate, gas fraction, age, and size.
\item We demonstrate in Section~\ref{sec:c4_lowz} the consistency of our result with the low-$z$ kinematics versus \lya\ study of \citet{Herenz2016}: a result that suggests the utility of \lya\ emission as a diagnostic of galaxy kinematics and other   properties over a wide range of redshifts.
\end{enumerate}

In Paper\,I in this series \citep{Foran2023} we report the photometric segregation of $z\sim2$ LBGs versus net \lya\ EW in rest-frame UV colour--magnitude space, and derive criteria for the selection of pure samples of LBGs with \lya\ dominant in absorption and \lya\ dominant in emission on the basis of optical broadband imaging alone.  Together with the analogous $z\sim3$ result of \citet{Cooke2009}, we have suggested the utility of this method to study a wide range of properties known to be associated with \lya\ (see Section~\ref{sec:c4_others}), in large samples and over large scales in datasets from current and future large-area and all-sky photometric surveys such as the Vera Rubin Observatory Legacy Survey of Space and Time \citep[LSST:][]{Ivezic2019}. 

Here we add nebular emission-line kinematics to the list of properties that might be studied by such an approach.  We propose a method by which the generalised kinematic type of large samples of LBGs might be determined on the basis of net \lya\ spectral types determined from broadband imaging, and their relation to other properties studied on large scales and at redshifts beyond the range accessible by current IFU spectrographs.  

The small size of the kinematic sample that meets the necessary selection criteria for this work (see Section~\ref{sec:c4_overview}) precludes a robust investigation of the relationship between net \lya\ EW and kinematics at fixed values of other galactic properties.  Nevertheless, these results (i) suggest the potential value of a holistic interpretation of galaxy evolution in terms of these many correlated properties, including kinematics and their relation to galaxy environment,  and (ii) provide motivation for a dedicated high-resolution AO observational campaign that specifically targets a larger, uniformly-selected and analysed sample that would inform the strength of all these relations, and enable application of multi-variate regression techniques to determine how they are related causally or otherwise (cf.\ the low-$z$ LARS study of \citet{Runnholm2020}).

Finally, in Section~\ref{sec:c4_implications} we speculate that the combination of our result linking net \lya\ EW and nebular emission-line kinematics with the known large-scale clustering behaviour of \lya-absorbing and \lya-emitting LBGs \citep{Cooke2013}, is evocative of an emergent bimodality of early galaxies that is consistent with a nascent morphology-density relation being observed at $z\sim2-3$ (Section~\ref{sec:c4_implications}).  

\bibliography{all_refs}


\end{document}